\documentclass[useAMS,usenatbib]{mn2e}
\usepackage{txfonts,graphicx,natbib}
\usepackage{longtable}
\bibpunct{(}{)}{;}{a}{}{,}
\newcommand{\Msun}{M$_{\odot}$}

\newcommand{\Fefs}{$^{56}$Fe}
\newcommand{\Cofs}{$^{56}$Co}
\newcommand{\Nifs}{$^{56}$Ni}

\newcommand{\DeltaB}{$\Delta m_{15}(B)$}
\newcommand{\DeltaV}{$\Delta m_{15}(V)$}
\newcommand{\DeltaBol}{$\Delta m_{15}(Bol)$}

\begin{document}

\title[ESC observations of SN 2005cf: Light Curves]{ESC observations
of SN 2005cf \\
I. Photometric Evolution of a Normal Type Ia Supernova} \author[Pastorello et al.]{A. Pastorello$^{1,2}$
\thanks{e--mail: pasto@MPA-Garching.MPG.DE}, 
S. Taubenberger$^{1}$, N. Elias--Rosa$^{1,3,4}$, P. A. Mazzali$^{1,5}$,
 G. Pignata$^{6}$, \and   E. Cappellaro$^{3}$,  G. Garavini$^{7}$, S. Nobili$^{7}$, 
G. C. Anupama$^{8}$, D. D. R. Bayliss$^{9}$, S. Benetti$^{3}$, \and 
F. Bufano$^{3,10}$,  N. K. Chakradhari$^{8}$, 
R. Kotak$^{11}$, A. Goobar$^{7}$,
H. Navasardyan$^{3}$, 
F. Patat$^{11}$, \and   D. K. Sahu$^{8}$, M. Salvo$^{9}$,
B. P. Schmidt$^{9}$,  V. Stanishev$^{7}$,
M. Turatto$^{3}$, 
and W. Hillebrandt$^{1}$\\
$^{1}$ Max-Planck-Institut f\"{u}r Astrophysik,
Karl-Schwarzschild-Str. 1, D-85741 Garching bei M\"{u}nchen, Germany\\
$^{2}$ Astrophysics Research Centre, School of Mathematics and Physics, Queen's
University Belfast, Belfast BT7 1NN, United Kingdom\\
$^{3}$ INAF Osservatorio Astronomico di Padova, Vicolo dell'Osservatorio 5, I-35122 Padova, Italy\\ 
$^{4}$ Universidad de La Laguna, Av. Astrof\'isico Francisco S\'anchez
s/n, E-38206 La Laguna, Tenerife, Spain\\
$^{5}$ INAF Osservatorio Astronomico di Trieste, Via Tiepolo 11, I-34131 Trieste, Italy\\
$^{6}$ Departamento de Astronom\'ia y Astrof\'isica, Pontificia Universidad Cat\'olica de Chile, Casilla 306, Santiago 22, Chile \\
$^{7}$ Department of Physics, Stockholm University, AlbaNova University Center, SE-10691 Stockholm, Sweden\\
$^{8}$ Indian Institute of Astrophysics, Koramangala, Bangalore 560 034, India\\
$^{9}$ Research School of Astronomy and Astrophysics, Australian
National University, Cotter Road, Weston Creek, ACT 2611, Australia\\
$^{10}$ Dipartimento di Astronomia, Universit\`a di Padova, Vicolo dell'Osservatorio 2, I-35122 Padova, Italy\\
$^{11}$ European Southern Observatory (ESO), Karl-Schwarzschild-Str. 2,
D-85748 Garching bei M\"{u}nchen, Germany\\
\\
}
\date{Accepted .....; Received ....; in original form ....}

\maketitle

\begin{abstract}
We present early--time optical and near--infrared photometry of supernova
(SN) 2005cf. The observations, spanning a period from about 12 days before to 
3 months after maximum, have been obtained through the coordination of  
observational efforts of various nodes of the European
Supernova Collaboration and including data obtained at the
2m Himalayan Chandra Telescope. From the observed light curve we deduce
that SN 2005cf is a fairly typical SN Ia with a post--maximum 
decline ($\Delta m_{15}(B)_{true}$ = 1.12) close to
the average value and a normal
luminosity of $M_{B,max}$ = --19.39$\pm$0.33. 
Models of the bolometric light curve suggest a synthesised  
$^{56}$Ni mass of about 0.7M$_\odot$.
The negligible host galaxy interstellar extinction and its proximity 
make SN 2005cf a good Type Ia supernova template.

\end{abstract}


\section{Introduction}  \label{intro}

Type Ia supernovae (SNe Ia) have been extensively studied in recent
years for their important cosmological implications.
They are considered to be powerful distance indicators because they combine
a high luminosity with relatively homogeneous physical properties.
Moreover, observations of high--redshift SNe Ia have provided the clue
for discovering the presence of a previously undetected cosmological component
with negative pressure, labelled ``dark energy'', and responsible of the accelerated
expansion of the Universe \citep[see][and references therein]{asti06}.

Thanks to the collection of a larger and larger compendium of new
data \citep[e.g.][]{hamu96,ries99a,jha06}, an unexpected variety in the observed
characteristics of SNe Ia has been shown to exist, and it is only using 
empirical relations between luminosity and distance--independent 
parameters, e.g. the shape of the light curve \citep[see e.g.][]{phil93}, 
that SNe Ia can be used as standardisable candles. 

Actually \citet{ben05} have recently shown that a 
one--parameter description of SNe Ia
does not account for the observed variety of these
objects.
In order to understand the physical reasons that cause
the intrinsic differences in Type Ia SNe properties, we need 
to improve the statistics by studying in detail a larger 
number of nearby objects.
There are considerable advantages in analysing nearby SNe:  
one can obtain higher signal--to--noise (S/N) data, and 
these SNe can be observed for a longer time after the explosion, 
providing more information on the evolution during the nebular phase.
Moreover, due to their proximity, the host galaxies
are frequently monitored
by automated professional SN searches and/or individual amateur astronomers. 
This significantly increases the probability of discovering very young
SNe Ia, allowing the study of these objects at the earlier phases after the explosion.

In order to constrain the explosion and progenitor models, 
excellent--quality data of a significant sample of nearby SNe Ia is
necessary.
To this end, a large consortium of groups, comprising both observational 
and modelling expertise has been formed (European Supernova Collaboration, ESC) 
as part of a European Research Training Network (RTN)\footnote{http://www.mpa-garching.mpg.de/$\sim$rtn/}.

To date, we obtained high--quality data for about 15 nearby SNe Ia.
Analyses of individual SNe include SN 2002bo \citep{ben04},
SN 2002er \citep{pig04,kot05},  SN 2002dj (Pignata et al., in preparation),
SN 2003cg \citep{nancy06}, SN 2003du \citep{sta06}, SN 2003gs
(Kotak et al., in preparation), SN 2003kf (Salvo et al., in preparation), SN 2004dt
(Altavilla et al., in preparation) and SN 2004eo \citep{pasto06}.
Statistical analysis of samples of SNe Ia, including those followed 
by the ESC were performed by \citet{ben05}, \citet{maz05}, \citet{hach06}.

The proximity of SN~2005cf and its discovery almost two weeks before
the $B$-band maximum (see below) made it an ideal target for the ESC.
Immediately following the discovery announcement, we started 
an intensive photometric and spectroscopic monitoring campaign, which covered 
the SN evolution over a period of about 100 days from the discovery.
This is the first of two papers where ESC data of SN~2005cf are presented. 
This work is devoted to study the early time optical and IR photometric observations of SN 2005cf,
while spectroscopic data will be presented in a forthcoming paper \citep{gara06}.

The layout of this paper is as follows:
in Sect. \ref{sect-obs} the
ESC observations of SN~2005cf will be presented, including 
 a description of the data reduction techniques. In Sect. \ref{sect-LC} 
the light curves of SN~2005cf will be displayed and analysed. 
In Sect. \ref{sect-param} we derive the main parameters of the SN 
using empirical relations from literature, while in 
Sect. \ref{sect-model} additional properties are inferred from light curve 
modelling. We conclude the paper with a summary (Sect. \ref{sect-disc}). 

\begin{figure}
 \resizebox{\hsize}{!}{\includegraphics{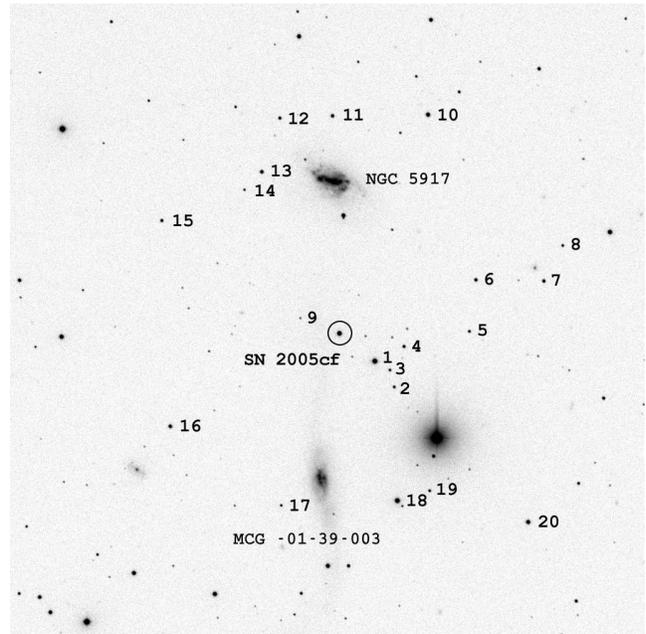}}
   \caption{SN 2005cf in MCG --01--39--003: 
$B$ band image obtained on July 1, 2005 with
the 2.2m Telescope of Calar Alto equipped with CAFOS.
The field of view is 9$\times$9 arcmin$^{2}$.
The sequence stars of Tab. \ref{seqstars} are
labelled with numbers. The galaxy below the SN in the image is MCG --01--39--003, the upper one
is NGC 5917. North is up, East is to the left.}
   \label{fig:SNfield}
\end{figure}

\begin{figure}
 \resizebox{\hsize}{!}{\includegraphics{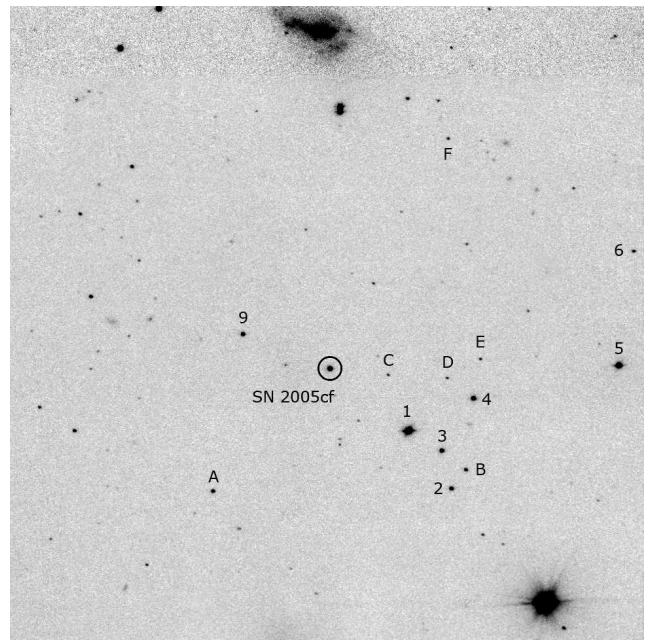}}
   \caption{SN 2005cf in MCG --01--39--003: 
$H$ band image obtained on July 14, 2005 with
the Telescopio Nazionale Galileo (TNG)  of La Palma (Canary Islands) 
equipped with NICS.
The field of view is 4.2$\times$4.2 arcmin$^{2}$.
The NIR sequence stars reported in Tab. \ref{IRseqstars} are
labelled.  North is up, East is to the left.}
   \label{fig:SNIRfield}
\end{figure}

\section{Observations} \label{sect-obs}

\subsection{SN 2005cf and the Host Galaxy} \label{sn_host}

SN~2005cf was located close to the tidal bridge between two galaxies.
It is known that the interaction between galaxies and/or galaxy
activity phenomena may enhance the rate of star formation. As a consequence
the rate of SNe in such galaxies is expected to increase.
Although this scenario should favour
mainly core--collapse SNe descending from short--lived progenitors,
it could also increase the number of progenitors of SNe Ia
with respect to the genuinely old stellar population
\citep{mdv94}.
Indeed \citet{smi81} found indications of enhanced production of SNe
of all types in interacting galaxy systems and \citet{nava01}
obtained a similar result in interacting pairs.
More recently \citet{mdv05} found evidence of an
enhanced rate of SNe Ia in radio--loud galaxies, probably due to repeated
episodes of interaction and/or merging.
However, in general, the location of SN explosions does not seem to coincide
with regions of strong interaction in the galaxies
\citep{nava01}, and the discovery of SNe in tidal tails remains an exceptional
event  \citep{pet95}.
This makes SN~2005cf a very interesting case.

SN~2005cf was discovered by H. Pugh and W. Li with the KAIT telescope
on May 28.36 UT when it was at magnitude 16.4 \citep{puck05}.
\citet{puck05} also report that nothing was visible on May 25.37 UT to
a limiting magnitude of 18.5.
The coordinates of SN 2005cf are $\alpha = 15^{h}21^{m}32\fs21$ and $\delta  = 
-07\degr24\arcmin47\farcs5$ (J2000). 
The object is located 15".7 West and 123" North of the centre 
of MCG --01--39--003 (Fig. \ref{fig:SNfield}), a peculiar S0 galaxy 
(source NED). 

SN~2005cf lies in proximity of a luminous bridge connecting MCG --01--39--003  
with the Sb galaxy MCG --01--39--002 (also known as NGC 5917, 
see Fig. \ref{fig:SNfield}). This makes
the association to one or the other galaxy uncertain. We will
assume that SN~2005cf exploded in  MCG --01--39--003, in agreement with 
\citet{puck05}, remarking that such assumption has no
significant effect on the overall SN properties.   
Basic information on SN 2005cf, its host galaxy and the interacting
companion is listed in Tab. \ref{gal_param}.

\citet{modj05} obtained a spectrum on May 31.22 UT with the Whipple Observatory 
1.5-m telescope (+ FAST) and classified the new object as a young 
(more than 10 days before maximum light) Type Ia SN.
This gave the main motivation for the activation of the follow--up 
campaign by the ESC.

\begin{table}
\begin{center}
\caption{Main parameters for SN 2005cf, the host galaxy MCG
--01--39--003 and the interacting companion NGC 5917.
\label{gal_param}}
\footnotesize
\begin{tabular}{ccc}\hline \hline
\multicolumn{3}{c}{SN 2005cf} \\ \hline 
$\alpha$ (J2000.0) & 15$^{h}$21$^{m}$32\fs21 &1 \\
$\delta$ (J2000.0) & -07$\degr$24$\arcmin$47$\farcs$.5 & 1\\
Offset Galaxy--SN$^{\ddag}$ & 15$^{\prime\prime}$.7W, 123$^{\prime\prime}$N &1 \\
SN Type & Ia & 2 \\
E$(B-V)_{host}$ & 0 & 3 \\
E$(B-V)_{Gal}$ & 0.097 & 4 \\ 
Discovery date (UT)& 2005 May 28.36 & 1 \\
Discovery JD & 2453518.86 & 1 \\
Discovery mag. & 16.4 & 1 \\
Predisc. limit epoch (UT)& 2005 May 25.37 & 1 \\
Predisc. limit mag. & 18.5 & 1 \\
JD($B_{max}$) & 2453534.0 & 3 \\
$B_{max}$ & 13.54 & 3 \\
$M_{B,max}(\mu)$ & $-$19.39  & 3 \\
$\Delta$m$_{15}(B)_{true}$ & 1.12 & 3 \\ 
$s^{-1}$ & 0.99 & 3 \\ 
$\Delta$$C_{12}$ & 0.355 & 3 \\ 
t$_r$ & 18.6   & 3 \\
M($^{56}$Ni) & 0.7M$_\odot$ & 3  \\ \hline \hline
\multicolumn{3}{c}{MCG --01--39--003} \\ \hline
PGC name & PGC 054817 & 5 \\
Galaxy type & S0 pec & 5 \\
$\alpha$ (J2000.0) & 15$^{h}$21$^{m}$33$\fs$29 & 5 \\
$\delta$ (J2000.0) & $-07\degr26\arcmin52\farcs38$ & 5 \\
$B_{tot}$ & 14.75 $\pm$ 0.43 & 6 \\
D & $1\arcmin4 \times 0\arcmin7$ & 5 \\
v$_{Vir}$ & 1977 km s$^{-1}$ & 6 \\
v$_{3k}$ & 2114 km s$^{-1}$ & 6 \\
$\mu$ & 32.51 &  3 \\
E$(B-V)_{Gal}$ & 0.098 & 4 \\ \hline \hline 
\multicolumn{3}{c}{MCG --01--39--002 (NGC 5917)} \\ \hline
PGC name & PGC 054809 & 5 \\
Galaxy type & Sb pec & 5 \\
$\alpha$ (J2000.0) & 15$^{h}$21$^{m}$32$\fs$57 & 5 \\
$\delta$ (J2000.0) & $-07\degr22\arcmin37\farcs8$ & 5 \\
$B_{tot}$ & 13.81 $\pm$ 0.50 & 6 \\
D & $1\arcmin5 \times 0\arcmin9$ & 5 \\
v$_{vir}$ & 1944 km s$^{-1}$ & 6 \\
v$_{3k}$ & 2080 km s$^{-1}$ & 6 \\
$\mu$ & 32.51 &  3 \\
E$(B-V)_{Gal}$ & 0.095 & 4 \\
\hline 
\end{tabular}

$^{\ddag}$ we assume MCG --01--39--003 as host galaxy.\\
1 = \protect\cite{puck05};
2 = \protect\cite{modj05};\\
3 = this paper;
4 = \protect\cite{schl98}; \\
5 = NED ({it http://nedwww.ipac.caltech.edu});\\
6 = LEDA ({it http://leda.univ-lyon1.fr}).
\end{center}
\end{table}

\subsection{ESC Observations} \label{instr_descr}

\begin{table*}
\caption{$U$, $B$, $V$, $R$, $I$ magnitudes of the sequence stars in the field of SN
2005cf. The errors reported in brackets are the r.m.s. of the
available measurements, obtained during photometric nights only.}
\centering
\label{seqstars}
\begin{tabular}{cccccc}
\hline\hline
Star & $U$ & $B$ & $V$ & $R$ & $I$ \\ \hline
 1 & 14.886 (0.018) &14.710 (0.008) &13.986 (0.007) &13.561 (0.007)&13.155 (0.006)\\ 
 2 & 18.243 (0.021) &18.125 (0.008) &17.308 (0.007) &16.867 (0.008)&16.453 (0.006)\\ 
 3 & 19.217 (0.015) &18.292 (0.009) &17.262 (0.007) &16.682 (0.008)&16.178 (0.006)\\ 
 4 & 18.080 (0.020) &17.548 (0.011) &16.661 (0.007) &16.164 (0.008)&15.705 (0.006)\\ 
 5 & 19.290 (0.013) &18.131 (0.009) &16.705 (0.005) &15.800 (0.005)&14.930 (0.007)\\ 
 6 &  &17.176 (0.008) &17.005 (0.008) &16.940 (0.012)&16.829 (0.007)\\ 
 7 & 17.314 (0.020) &17.172 (0.010) &16.350 (0.006) &15.895 (0.008)&15.436 (0.006)\\ 
 8 & 19.128 (0.022) &18.207 (0.009) &17.152 (0.008) &16.552 (0.006)&16.035 (0.007)\\ 
 9 & 20.984 (0.059) &19.776 (0.010) &18.490 (0.012) &17.605 (0.007)&16.795 (0.006)\\ 
10 & 15.865 (0.009) &15.827 (0.011) &15.169 (0.007) &14.777 (0.005)&14.423 (0.006)\\ 
11 & 16.917 (0.010) &16.929 (0.006) &16.264 (0.007) &15.883 (0.006)&15.523 (0.010)\\ 
12 & 17.475 (0.011) &17.503 (0.008) &16.837 (0.007) &16.458 (0.007)&16.109 (0.005)\\ 
13 & 17.668 (0.016) &16.688 (0.008) &15.662 (0.007) &15.066 (0.006)&14.573 (0.009)\\ 
14 & 19.939 (0.018) &18.879 (0.009) &17.271 (0.010) &16.300 (0.008)&15.167 (0.010)\\ 
15 & 17.627 (0.011) &17.398 (0.009) &16.543 (0.009) &16.058 (0.008)&15.602 (0.005)\\ 
16 & 16.446 (0.011) &16.401 (0.010) &15.700 (0.007) &15.288 (0.006)&14.909 (0.008)\\ 
17 & 18.181 (0.016) &18.456 (0.009) &17.863 (0.007) &17.487 (0.009)&17.153 (0.007)\\ 
18 & 15.228 (0.019) &14.717 (0.008) &13.833 (0.006) &13.298 (0.006)&12.786 (0.008)\\ 
19 & 17.808 (0.013) &17.952 (0.008) &17.397 (0.009) &17.031 (0.007)&16.695 (0.007)\\ 
20 & 16.556 (0.022) &15.813 (0.006) &14.801 (0.005) &14.201 (0.005)&13.676 (0.007)\\ \hline
\end{tabular}
\end{table*}

\begin{table}
\caption{$J$, $H$, $K'$ magnitudes of the sequence stars in the field of SN
2005cf, with assigned errors.}
\centering
\label{IRseqstars}
\begin{tabular}{cccc}
\hline\hline
Star & $J$ & $H$ & $K'$ \\ \hline
 1 & 12.54 (0.01) & 12.15 (0.02) & 12.21 (0.01) \\
 2 & 15.91 (0.01) & 15.46 (0.02) & 15.53 (0.03) \\
 3 & 15.49 (0.01) & 14.97 (0.01) & 14.99 (0.02) \\
 4 & 15.10 (0.02) & 14.61 (0.02) & 14.60 (0.01) \\
 5 & 13.84 (0.08) & 13.22 (0.01) & 13.25 (0.03) \\
 6 & 16.74 (0.01) & 16.50 (0.01) &   \\
 9 & 15.80 (0.01) & 15.07 (0.03) & 15.01 (0.02) \\
 A & 16.56 (0.03) & 15.88 (0.04) & 15.85 (0.04) \\
 B & 16.85 (0.03) & 16.27 (0.03) & 16.12 (0.01) \\
 C & 17.70 (0.03) & 17.31 (0.02) & 17.34 (0.01) \\
 D & 17.77 (0.03) & 17.36 (0.04) & 17.40 (0.01) \\
 E & 17.62 (0.01) & 17.26 (0.02) & 17.27 (0.07) \\
 F & 17.39 (0.01) & 17.26 (0.07) &  \\ \hline
\end{tabular}
\end{table}

We have obtained more than 360 optical data points, covering about 60
nights, from about 12 days before the
$B$ band maximum to approximately 3 months after. In
addition,  near--infrared (NIR) observations have been performed in 5 selected
epochs. Observations at late phases will be presented in a forthcoming paper.

During the follow--up, 8 different instruments have been used for the
optical and 2 for the NIR observations:
\begin{itemize}
\item the 40--inch Telescope at the Siding Spring Observatory
(Australia) with a Wide Field Camera (eight
2048$\times$4096 CCDs, with pixel scale of 0.375 arcsec pixel$^{-1}$)
and standard broad band Bessell filters $B$, $V$, $R$, $I$;
\item the 3.58m Italian Telescopio Nazionale Galileo (TNG) at the Observatorio de los Muchachos
in La Palma (Canary Islands, Spain),
equipped with DOLORES and a Loral thinned and back--illuminated 
2048$\times$2048 detector, with scale 0.275 arcsec pixel$^{-1}$, 
yielding a field of view of about 9.4$\times$9.4 arcmin$^2$. We used
the $U$, $B$, $V$ Johnson and $R$, $I$ Cousins filters (with TNG identification
numbers 1, 10, 11, 12, 13, respectively);
\item the 2.5m Nordic Optical Telescope  in La Palma equipped with ALFOSC
(with an E2V 2048$\times$2048 CCD of 0.19 arcsec pixel$^{-1}$) and a
set of $U$, $B$, $V$, $R$ Bessell filters (with NOT identification numbers 7, 74, 75, 76,
respectively) and an interference $i$ band filter (number 12);
\item the 2.3m Telescope in Siding Spring, equipped
with the E2V 2048$\times$2048 imager, with pixel scale of 0.19
arcsec pixel$^{-1}$ and standard broad band Bessell filters $U$, $B$, $V$, $R$, $I$;
\item the Mercator 1.2m Telescope in La Palma, equipped with a 2048$\times$2048
CCD camera (MEROPE) having a field of view of 6.5$\times$6.5 arcmin$^2$ and a resolution of about
0.19 arcsec pixel$^{-1}$. We used $U$, $B$, $V$, $R$, $I$ filters
(with identification codes UG, BG, VG, RG and IC, respectively);
\item the 2.2m Telescope of Calar Alto (Spain) equipped with CAFOS and
 a SITe 2048$\times$2048 CCD, 0.53 arcsec pixel$^{-1}$; the filters 
available were the $U$, $B$, $V$, $R$, $I$ Johnson (labelled as 370/47b, 451/73, 534/97b,
641/158, 850/150b, respectively);
\item the Copernico 1.82m Telescope of Mt. Ekar (Asiago, Italy);
equipped with AFOSC, a TEKTRONIX 1024$\times$1024 thinned CCD (0.47
arcsec pixel$^{-1}$) and a set of $B$, $V$, $R$ Bessell and the $i$ Gunn filters;
\item the ESO/MPI 2.2m Telescope in La Silla (Chile), with a wide
field mosaic of eight 2048$\times$4096 CCDs (0.24 arcsec
pixel$^{-1}$) and a total field of view of 34$\times$33 arcmin$^2$. We
used the broad band filters $U$ (labelled as ESO877), $B$ (ESO878), $V$ (ESO843),
Cousins $Rc$ (ESO844), and EIS $I$ (ESO879);
\item the TNG equipped with the Near Infrared Camera Spectrometer
(NICS) with a HgCdTe Hawaii 1024$\times$1024 array (field of view
4.2$\times$4.2 arcmin$^2$, scale 0.25 arcsec pixel$^{-1}$); the filters used in the observations were  $J$, $H$, $K'$.
\item the 3.5m Telescope in Calar Alto with Omega--Cass, having a Rockwell
1024$\times$1024 HgCdTe Hawaii array with pixel scale 0.2 arcsec
pixel$^{-1}$ and $J$, $H$, $K'$ filters.
\end{itemize}

\begin{figure}
\resizebox{\hsize}{!}{\includegraphics{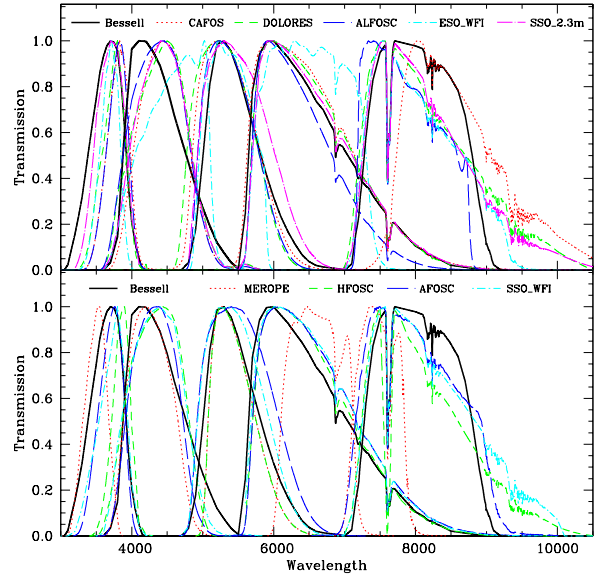}}
\caption{Comparison between the different instrumental $U$, $B$, $V$, $R$ and $I$
transmission curves, normalised to the peak transmission, and the standard Johnson--Cousins functions.
The instrumental transmission curves were reconstructed taking into account the filter transmission,
the CCD quantum efficiency, the reflectivity of the mirrors, the atmospheric transmission and, when available, 
the lenses throughput.
\label{filters} }
\end{figure}

In this paper we also included the data from Anupama et al. (in preparation)
obtained using the 2m Himalayan Chandra Telescope (HCT) of the Indian
Astronomical Observatory (IAO), Hanle (India), equipped with 
the Himalaya Faint Object Spectrograph Camera (HFOSC), with a SITe 
 2048$\times$4096 CCD (pixel scale 0.296 arcsec pixel$^{-1}$),
with a central region  (2048$\times$2048 pixels) 
used for imaging and covering a field of view of 10$\times$10
arcmin$^2$. Standard Bessell $U$, $B$, $V$, $R$, $I$ filters were used.

The $U$, $B$, $V$, $R$, $I$ transmission curves for all optical instrumental 
configurations used
during the follow--up of SN~2005cf are shown in Fig. \ref{filters}, 
and compared with the standard
Johnson--Cousins passbands \citep{bess90}.

\begin{table*}
  \caption{Comparison between synthetic and photometric colour terms. The colour terms for HFOSC
are provided by the observatory without associated errors.
For SSO--40inch Telescope + WFI, we have only two colour term measurements, therefore the
statistic is too poor to compute reliable errors.
} \label{synt_photo}
  \scriptsize
  \label{symbols}
  \begin{tabular}{@{}lcccccccccc}
  \hline
\multicolumn{11}{l}{Instrument ~~~~~~~~~~~~~~~~~  $U$~$(U-B)$ ~~~~~~~~~~~~~~~~~~~~~~~~~~~~~~~~~~~~
$B$~$(B-V)$ ~~~~~~~~~~~~~~~~~~~~~~~~~~~~~~~~~~~  $V$~$(B-V)$  ~~~~~~~~~~~~~~~~~~~~~~~~~~~~~~~~~~~~ $R$~$(V-R)$  ~~~~~~~~~~~~~~~~~~~~~~~~~~~~~~~~~~~~
$I$~$(R-I)$}\\
  & syn & ph & syn & ph & syn & ph & syn & ph & syn & ph \\
\hline
40inch+WFI &  &  & $-$0.022 & 0.005  & $-$0.002 & 0.033 & 0.054 & 0.054 & 0.042 & $-$0.019 \\
DOLORES  & 0.059$\star$    &  0.105 $\pm$ 0.047  & 0.080   &0.064 $\pm$ 0.018  &$-$0.101 & $-$0.120 $\pm$ 0.033 &  0.031 &  0.022$\pm$ 0.037 &  0.025 &  0.023 $\pm$ 0.017 \\
ALFOSC &  0.093$\star$   &  0.122 $\pm$ 0.010  & 0.023   &  0.044 $\pm$ 0.012  &$-$0.048 & $-$0.049 $\pm$ 0.021 & $-$0.074 &  $-$0.098 $\pm$ 0.024 &  $-$0.086 &  $-$0.068 $\pm$ 0.033 \\
2.3m+imager&  0.048    &  0.094 $\pm$ 0.033  & 0.081   &  0.014 $\pm$ 0.024  &  0.026   &  0.027 $\pm$ 0.009  & 0.035   &  0.040 $\pm$ 0.016 & 0.028 &  0.002 $\pm$ 0.016 \\
Merope & $-$0.089 &  $-$0.105 $\pm$ 0.008 & $-$0.168  &  $-$0.130 $\pm$ 0.014 & $-$0.002 &  $-$0.001 $\pm$ 0.011 & ~~0.126 & ~~0.134 $\pm$ 0.022 & $-$0.284  & $-$0.332 $\pm$ 0.026  \\ 
CAFOS  & 0.121$\star$   & 0.167 $\pm$ 0.016 &  0.107  &  0.115  $\pm$ 0.005  & $-$0.061 & $-$0.048 $\pm$ 0.005 &  $-$0.008 & $-$0.015 $\pm$ 0.017 &  0.219 &  0.256 $\pm$ 0.036  \\
AFOSC &  &  & $-$0.068$\star$ & $-$0.030 $\pm$ 0.013 & 0.041 &0.047 $\pm$ 0.010 & 0.066 & 0.052  $\pm$ 0.033 & $-$0.031 &  $-$0.044 $\pm$ 0.036 \\
2.2m+WFI & $-$0.021  & $-$0.070 $\pm$ 0.036 & 0.249 & 0.244 $\pm$ 0.024 & $-$0.067 & $-$0.067 $\pm$ 0.013 & 0.015 & 0.000 $\pm$ 0.022 &  $-$0.010 &0.006 $\pm$ 0.022 \\
HFOSC & 0.155 & 0.188 & $-$0.028  &  $-$0.049 & 0.045 & 0.047 & 0.038& 0.065 &  0.018 & 0.017 \\

\hline
\end{tabular}

$\star$ ~The passband's blue cutoff was modified
\end{table*}  

\subsection{Data Reduction}

The reduction of the optical photometry was performed using standard
IRAF\footnote{IRAF is distributed by the National Optical Astronomy
Observatories, which are operated by the Association of Universities
for Research in Astronomy, Inc., under cooperative agreement with the National
Science Foundation.} tasks. The first reduction steps included 
bias, overscan and flat field corrections, and the trimming
of the images using the IRAF package {\it CCDRED}. 

The pre--reduction of the NIR images was slightly more laborious, 
as it required a few additional steps.
Due to the high luminosity of the night sky in the NIR, we needed to 
remove the sky contribution from the target images, by creating
``clean'' sky images. This was done 
by median--combining a number of dithered science frames. 
The resulting sky template image was then subtracted from the target images.
Most of our data were obtained with several short--exposure dithered
frames, which had to be spatially registered and then combined in 
order to improve the S/N.

The NICS images required particular treatment, as they needed also to be
corrected for the cross talking effect (i.e. a signal which was detected in one quadrant
produced negative ghost images in the other 3 quadrants) and for the
distortion of the NICS optics. These corrections were performed using 
a pipeline,  {\it SNAP}\footnote{{\it http://www.arcetri.astro.it/$\sim$filippo/snap/}}, 
available at TNG for the reduction of images obtained using
NICS.

Instrumental magnitudes of SN~2005cf were determined with the
point--spread function (PSF) fitting technique, performed using the
{\it SNOoPY}\footnote{{\it SNOoPY} is a package originally designed by F. Patat, implemented
in IRAF by E. Cappellaro and based on {\it DAOPHOT}.} package. Since SN 2005cf
is a very bright and isolated object, the subtraction of the host
galaxy template is not required, and the PSF fitting technique
provides excellent results.

In order to transform instrumental magnitudes into the standard
photometric system, first--order colour corrections were
applied, using colour terms derived from observations of several photometric
standard fields \citep{land92}. 
The photometric zeropoints were finally determined for all nights 
by comparing magnitudes of a local sequence of stars in the vicinity of the
host galaxy (cf. Fig. \ref{fig:SNfield}) 
to the average estimates obtained during some photometric nights.
The average magnitudes for the sequence stars in the field of
SN~2005cf are  reported in Tab. \ref{seqstars}. 

The HFOSC data from Anupama et al. have been checked
comparing the stars in common to both local sequences.
In order to calibrate their SN magnitudes onto our sequence, we applied
additive zeropoint shifts (smaller than 0.05 mags), 
slightly corrected for the colour terms of the instrumental configuration of HCT. 

In analogy to optical observations, NIR photometry was computed using
different standard fields of the Arnica catalogue \citep{hunt98} and finally calibrated 
using a number of local standards in the field of SN~2005cf (Fig. \ref{fig:SNIRfield}).
The $J$, $H$, $K'$ magnitudes of the IR local standards are reported in
Tab. \ref{IRseqstars}.

\section{Light Curves of SN~2005cf} \label{sect-LC}

\subsection{S--Correction to the Optical Light Curves}

The optical photometry of SN~2005cf, as derived from comparison with the 
Landolt's standard fields and our local sequence stars only (see below),
shows a disturbing scatter in the magnitudes obtained
using different instrumental configurations.
 This was due to the combination of the difference between the instrumental
photometric system (see also Fig. \ref{filters}) and the non--thermal SN spectrum.
In order to remove these systematic errors we used a technique, presented in
\citet{max02}, called S--correction. To compute the corrections, one
 first needs to determine the instrumental passband
$S(\lambda)$, defined as:

\begin{equation}
S(\lambda)=F(\lambda)\cdot QE(\lambda) \cdot A(\lambda) \cdot M(\lambda) \cdot L(\lambda)
\end{equation}

\noindent where $F(\lambda)$ is the filter transmission function, $QE(\lambda)$
is the detector quantum efficiency, $A(\lambda)$ is the continuum atmospheric
transmission profile, $M(\lambda)$ is the mirror reflectivity function and 
$L(\lambda)$ is the
lens throughput. Information on instruments, detectors and filters
used during the follow--up of SN~2005cf is given in Sect. \ref{instr_descr}. 
In order to derive the atmospheric transmission profile of Calar Alto and La Palma, we
made use of the information reported in
\citet{atm_caha} and \citet{atm_lapalma}, respectively, while for the
La Silla site we used the CTIO transmission curve available in IRAF. 
For Asiago--Ekar, the Siding Spring Observatory and the Indian Astronomical
Observatory we obtained $A(\lambda)$ by adapting the standard atmospheric
model proposed by \citet{walker} in order to match the average broad band
absorption coefficients of these sites. 
Finally, $M(\lambda)$ was obtained using a standard aluminum reflectivity curve multiplied
by the number of reflections in a given instrumental configuration, while
$L(\lambda)$ was estimated for DOLORES and WFI only.
For all the other instrumental configurations, we assumed that
$L(\lambda)$ was constant across the whole spectral range.
This approximation, together with a rapid variability both in the 
CCD quantum efficiency and in the atmosphere's transmission curve 
at the blue wavelengths, are 
probably the reasons why the $U$ reconstructed passbands do not match
the observed ones (see Tab. \ref{synt_photo}).

\begin{figure}
\resizebox{\hsize}{!}{\includegraphics{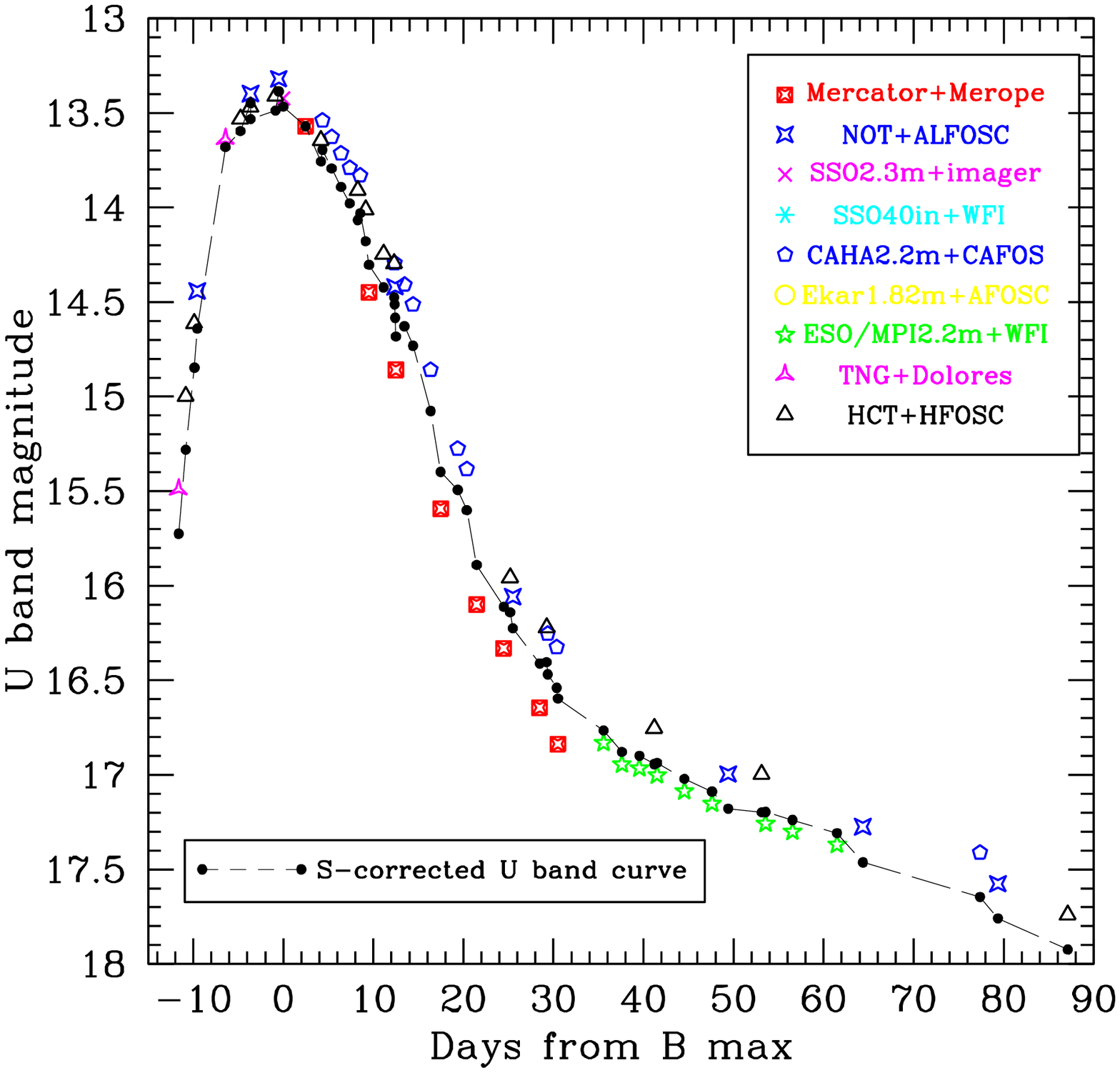}}
\resizebox{\hsize}{!}{\includegraphics{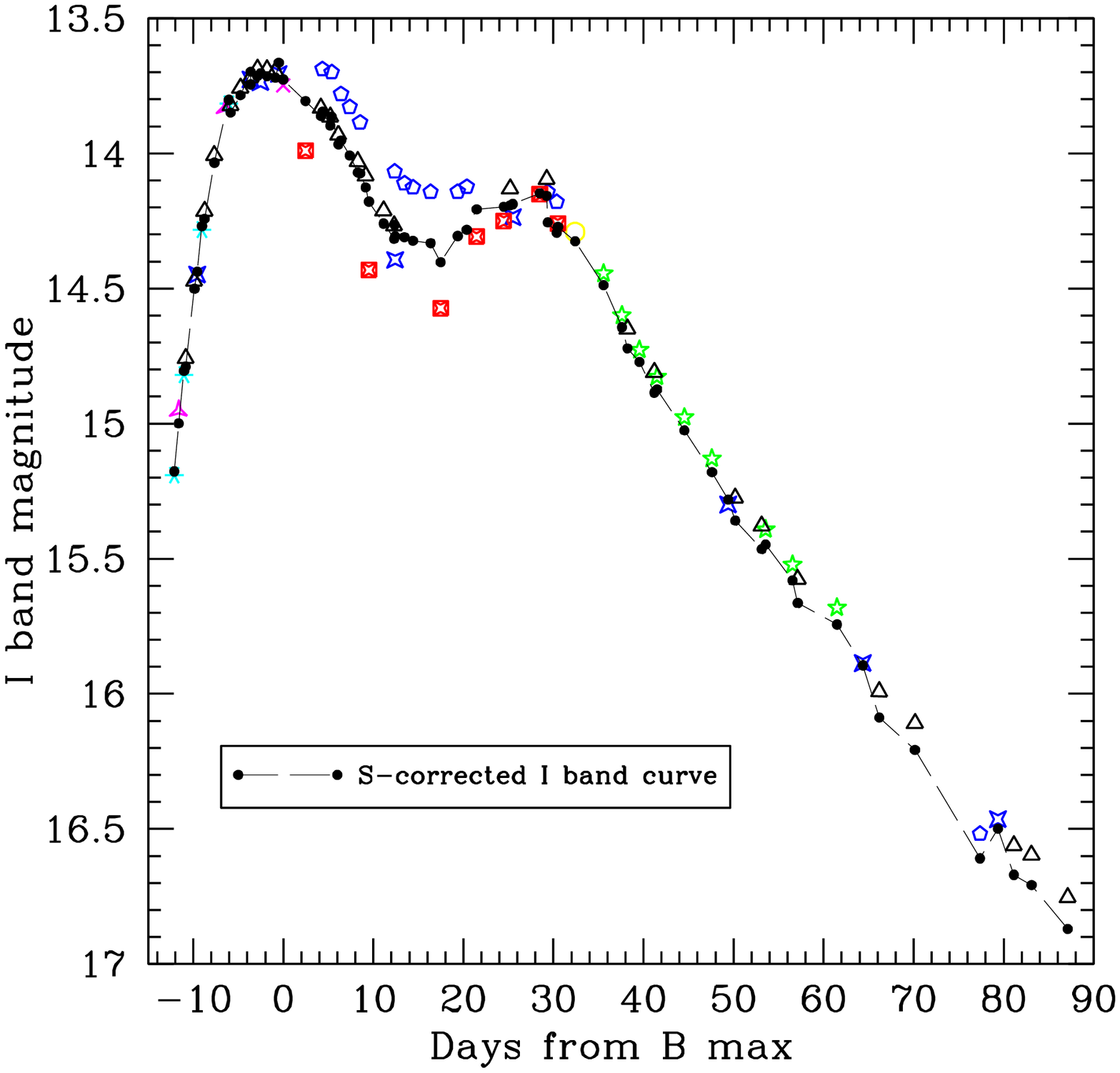}}
\caption{Top: comparison between the original $U$ band light curve of
SN~2005cf and the S--corrected one. Bottom: the same, but for the $I$ band. Original data
points obtained using different instruments are plotted with different
symbols. The large scatter is due to the diversity of the 
$U$ and $I$  filters available at the various telescopes. 
The final, S--corrected $U$ and $I$ band photometry is represented by filled 
circles, connected with a dashed line.
\label{corr_uncorr} }
\end{figure}

In order to check the match between the modelled
passbands and the real ones, and to
calculate the instrumental zero points for all configurations, 
we followed the same approach as \citet{pig04}, updating to 
the new set of spectro--photometric standard stars
from \citet{max05}, which span a range in colour
larger than that provided by previous works \citep[e.g.][]{ham94}.  
In Tab. \ref{synt_photo} we report the comparison
between the colour terms\footnote{The colour term is the coefficent $B$ of the equation 
$m_{\lambda 1} = m^{ST}_{\lambda 1} + A + B \times (m_{\lambda 2}- 
m_{\lambda 1})$ which is used to calibrate instrumental magnitude 
$m^{ST}_{\lambda 1}$ to the standard system $m_{\lambda 1}$.} 
computed via the synthetic photometry and those
determined through the observation of a number of standard fields. For the
latter, the estimates and their associated errors (see
Tab. \ref{synt_photo}) were computed using a 
3$\sigma$--clipped average. 
In a few cases, when the difference between the synthetic colour term and 
photometric colour term was larger than 3$\sigma$, the passbands were 
adjusted.

\begin{figure}
 \resizebox{\hsize}{!}{\includegraphics{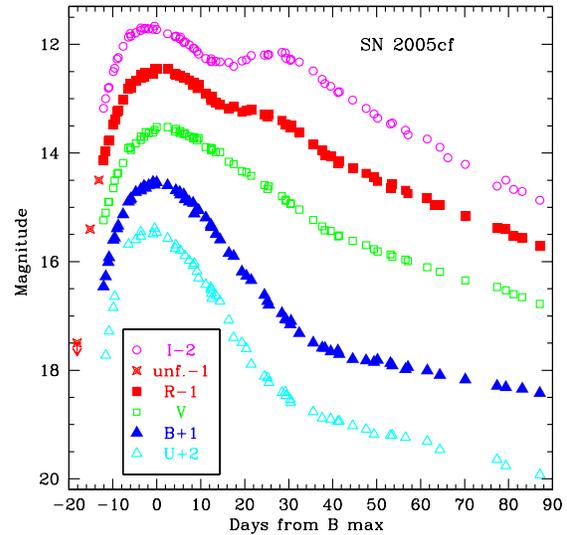}}
\caption{$U$, $B$, $V$, $R$, $I$ light curves of SN 2005cf, including the unfiltered 
measurements from IAU Circ. 8534 (asterisk). 
 The phase is measured in days from the $B$ band maximum.}
\label{fig:lightcurves}
\end{figure}

For each instrumental configuration, the S--correction measurements at
different epochs were fitted by a
third-- or fourth--order polynomial, as in \citet{pig04}, and
the r.m.s. deviations of the data points from the fitted law provided an
estimate of the errors due to the correction itself.
The correction applied to the photometry of SN~2005cf turned out
to be effective
because of the detailed characterisation of the photometric properties of
most of the instruments used by the ESC \citep{pig04} and
the excellent spectral sequence available for this object
\citep{gara06}.
Note that, however, since most spectra of SN~2005cf had not adequate
coverage in the region below 3500\AA, we had to resort to spectra
of SN 1994D in order to estimate the S--correction for 
the $U$ band.

The original (i.e. non S--corrected) optical photometry for SN~2005cf
is reported in Tab. \ref{SN_mags} (columns 3 to 7). 
The corrections to be applied to the original magnitudes
 are also reported in Tab. \ref{SN_mags} (columns 8 to 12).
The differences are in general quite small, especially in the $B$, $V$, $R$
bands, and they are significant only for some specific instrumental configurations 
(sometimes of 0.1--0.2 magnitudes, see e.g. the $I$ filters of the
Mercator Telescope + Merope, the 2.2m Calar Alto Telescope + CAFOS and
 the Himalayan Chandra Telescope + HFOSC, 
or the $B$ filter mounted at the 2.2m ESO/MPI Telescope + WFI). 
On the contrary, the $U$ band correction is large for most
instrumental configurations. As shown in Fig. \ref{filters} 
(see also Stanishev et al. 2006),  
this is because the sensitivity curves of the $U$ filters available at the 
various telescopes are significantly different
(being often shifted to redder wavelengths) compared to 
the standard $U$ Bessell passband.

\newpage

\onecolumn
\begin{center}
\begin{longtable}{ccccccccccccc} 
\caption{Original $U$, $B$, $V$, $R$, $I$ magnitudes of SN
2005cf (columns 3 to 7) and S--corrections (columns 8 to 12) 
 to be {\it added} to the original magnitudes to obtain the final, S--corrected
optical magnitudes of SN~2005cf. }\\ \hline\hline 
\multicolumn{1}{c}{Date} & \multicolumn{1}{c}{~~~~~~~JD} & \multicolumn{5}{c}{~~~~~~~~~~~~~~~~~~~~~~~~Original Magnitude} &
\multicolumn{5}{c}{~~~~~~~~~~~~~~~~~~~~~~~~~~S-correction} & \multicolumn{1}{c}{S}  \\ 
\label{SN_mags}
 & (+2400000) & $U$ & $B$  & $V$ & $R$& $I$ & $\Delta U$& $\Delta B$& $\Delta V$& $\Delta R$& $\Delta I$& \\ \hline
 31/5 & 53521.90&         & 15.493  & 15.243  & 15.142  & 15.191 &       & -0.031& -0.005& -0.004& -0.014 & 1 \\
 31/5 & 53522.38& 15.490  & 15.299  & 15.127  & 14.973  & 14.952 &  0.235& -0.027& -0.030&  0.003&  0.047 & 2 \\ 
 01/6 & 53522.97&         & 15.044  & 14.902  & 14.785  & 14.819 &       & -0.032& -0.004& -0.005& -0.014 & 1 \\ 
 01/6 & 53523.16& 14.999  & 14.934  & 14.913  & 14.769  & 14.759 &  0.282& -0.010& -0.011&  0.002&  0.030 & H \\ 
 02/6 & 53524.13& 14.613  & 14.609  & 14.658  & 14.466  & 14.472 &  0.234& -0.009& -0.009&  0.001&  0.029 & H  \\ 
 02/6 & 53524.44& 14.443  & 14.582  & 14.514  & 14.377  & 14.449 &  0.198& -0.021& -0.007&  0.006& -0.011 & 3 \\ 
 03/6 & 53524.97&         & 14.425  & 14.382  & 14.241  & 14.282 &       & -0.033& -0.003& -0.005& -0.013 & 1 \\ 
 03/6 & 53525.25&         & 14.348  & 14.373  & 14.224  & 14.214 &       & -0.009& -0.006&   0   &  0.028 &H \\ 
 04/6 & 53526.35&         & 14.139  & 14.189  & 14.009  & 14.007 &       & -0.008& -0.004& -0.002&  0.028 &H \\
 06/6 & 53527.57& 13.638  & 13.924  & 13.943  & 13.823  & 13.831 &  0.043& -0.026& -0.034& -0.001&  0.035 & 2 \\ 
 06/6 & 53527.95&         & 13.896  & 13.896  & 13.736  & 13.816 &       & -0.034& -0.002& -0.006& -0.014 & 1 \\ 
 06/6 & 53528.14&         & 13.848  & 13.944  & 13.792  & 13.822 &       & -0.007& -0.001& -0.004&  0.027 & H \\
 07/6 & 53529.24& 13.533  & 13.730  & 13.822  & 13.680  & 13.758 &  0.063& -0.006&  0.002& -0.005&  0.027 & H \\
 08/6 & 53530.38& 13.399  & 13.712  & 13.710  & 13.544  & 13.726 &  0.050& -0.018& -0.012&  0.012& -0.029 & 3 \\ 
 08/6 & 53530.38& 13.470  & 13.676  & 13.753  & 13.617  & 13.719 &  0.063& -0.006&  0.003& -0.007&  0.026 & H \\ 
 09/6 & 53531.12&         & 13.614  & 13.695  & 13.597  & 13.688 &       & -0.005&  0.004& -0.007&  0.025 & H \\ 
 09/6 & 53531.48&         & 13.659  & 13.629  &         & 13.737 &       & -0.017& -0.013&       & -0.034 & 3 \\ 
 10/6 & 53531.52&         &         &         & 13.501  &        &       &       & 	 &  0.013& 	  & 3 \\ 
 10/6 & 53532.19&         & 13.601  & 13.628  & 13.553  & 13.689 &       & -0.004&  0.004& -0.008&  0.025 & H  \\
 11/6 & 53533.13& 13.410  & 13.539  & 13.618  & 13.541  & 13.696 &  0.078& -0.004&  0.004& -0.009&  0.024 & H \\
 12/6 & 53533.50& 13.320  & 13.602  & 13.592  & 13.432  & 13.706 &  0.067& -0.015& -0.012&  0.015& -0.042 & 3 \\ 
 12/6 & 53533.99& 13.423  & 13.568  & 13.525  & 13.462  & 13.748 &  0.044& -0.029& -0.002& -0.007& -0.022 & 4 \\ 
 14/6 & 53536.47& 13.570  & 13.602  & 13.501  & 13.486  & 13.990 &  0.001& -0.008&  0.022& -0.041& -0.184 & 5 \\ 
 16/6 & 53538.19& 13.645  &         & 13.606  & 13.558  & 13.831 &  0.112&       & -0.005& -0.007&  0.031 & H \\
 16/6 & 53538.36& 13.540  & 13.725  & 13.581  & 13.562  & 13.688 &  0.155& -0.036& -0.023&  0.004&  0.158 & 6 \\ 
 17/6 & 53539.21&         & 13.723  & 13.641  & 13.583  & 13.864 &       & -0.003& -0.008& -0.007&  0.033 & H \\
 17/6 & 53539.37& 13.628  & 13.798  & 13.616  & 13.567  & 13.700 &  0.166& -0.034& -0.022&  0.004&  0.164 & 6 \\ 
 18/6 & 53540.13&         & 13.770  & 13.664  & 13.625  & 13.931 &       & -0.003& -0.010& -0.006&  0.036 & H \\
 18/6 & 53540.41& 13.714  & 13.887  & 13.656  & 13.607  & 13.780 &  0.178& -0.032& -0.021&  0.004&  0.172 & 6 \\ 
 19/6 & 53541.37& 13.790  & 13.929  & 13.713  & 13.653  & 13.828 &  0.189& -0.030& -0.020&  0.004&  0.180 & 6 \\ 
 20/6 & 53542.26& 13.909  & 13.925  & 13.743  & 13.742  & 14.030 &  0.158& -0.004& -0.016& -0.003&  0.041 & H \\
 21/6 & 53542.53& 13.831  & 14.144  & 13.724  & 13.695  & 13.885 &  0.201& -0.029& -0.019&  0.004&  0.189 & 6 \\ 
 21/6 & 53543.14& 14.013  & 14.050  & 13.781  & 13.814  & 14.083 &  0.165& -0.005& -0.018& -0.001&  0.043 & H \\
 22/6 & 53543.51& 14.450  & 14.065  & 13.694  & 13.764  & 14.432 & -0.146& -0.012&  0.033& -0.028& -0.254 & 5 \\ 
 23/6 & 53545.13& 14.246  & 14.197  & 13.915  & 13.906  & 14.212 &  0.177& -0.006& -0.023&  0.001&  0.047 & H \\
 24/6 & 53546.30& 14.296  & 14.330  & 13.974  & 14.031  & 14.267 &  0.179& -0.007& -0.025&  0.002&  0.049 & H \\ 
 24/6 & 53546.36& 14.294  & 14.386  & 13.921  & 13.974  & 14.067 &  0.219& -0.028& -0.012&  0.004&  0.198 & 6 \\ 
 24/6 & 53546.43& 14.420  & 14.368  & 13.937  & 13.950  & 14.394 &  0.163& -0.007&  0.003&  0.022& -0.087 & 3 \\  
 25/6 & 53546.50& 14.859  & 14.376  & 13.879  & 13.986  &        & -0.177& -0.014&  0.036& -0.018&        & 5 \\ 
 25/6 & 53547.45& 14.409  & 14.523  & 13.996  & 14.060  & 14.111 &  0.219& -0.028& -0.010&  0.004& -0.199 & 6 \\ 
 26/6 & 53548.40& 14.512 & 14.626  & 13.990  & 14.105  & 14.126   &  0.219& -0.029&-0.009&  0.005&  0.197 & 6 \\ 
 28/6 & 53550.37& 14.858 & 14.877  & 14.164  & 14.170  & 14.143   &  0.219& -0.033&-0.005&  0.005&  0.189 & 6 \\ 
 29/6 & 53551.48& 15.593 & 14.919  & 14.169  & 14.143  & 14.573   & -0.195& -0.019& 0.036& -0.005& -0.170 & 5 \\ 
 01/7 & 53553.37& 15.275 & 15.224  & 14.333  & 14.240  & 14.142   &  0.218& -0.040& 0.001&  0.006&  0.164 & 6 \\ 
 02/7 & 53554.37& 15.383 & 15.307  & 14.356  & 14.219  & 14.124   &  0.218& -0.042& 0.003&  0.007&  0.159 & 6 \\ 
 03/7 & 53555.47& 16.099 & 15.364  & 14.385  & 14.196  & 14.308   & -0.209& -0.024& 0.034&  0.005& -0.101 & 5 \\ 
 06/7 & 53558.48& 16.331 & 15.640  & 14.558  & 14.287  & 14.250   & -0.220& -0.024& 0.032&  0.004& -0.053 & 5 \\ 
 07/7 & 53559.19& 15.957 & 15.751  & 14.680  & 14.331  & 14.130   &  0.183& -0.025&-0.020& -0.011&  0.062 & H \\
 07/7 & 53559.49& 16.059 & 15.812  & 14.593  & 14.280  & 14.237   &  0.166& -0.015& 0.021&  0.007& -0.049 & 3 \\ 
 10/7 & 53562.48& 16.645 & 15.972  & 14.763  & 14.402  & 14.151   & -0.234& -0.020& 0.031& -0.001& -0.004 & 5 \\ 
 11/7 & 53563.24& 16.220 & 16.079  & 14.889  & 14.495  & 14.095   &  0.184& -0.026&-0.015& -0.018&  0.062 & H \\
 11/7 & 53563.37& 16.253 & 16.135  & 14.850  & 14.459  & 14.143   &  0.216& -0.067& 0.011&  0.010&  0.113 & 6 \\ 
 12/7 & 53564.37& 16.324 & 16.226  & 14.923  & 14.515  & 14.180   &  0.215& -0.068& 0.011&  0.010&  0.114 & 6 \\ 
 \\
\caption{continued.}\\ 
\hline\hline 
\multicolumn{1}{c}{Date} & \multicolumn{1}{c}{~~~~~~~JD} & \multicolumn{5}{c}{~~~~~~~~~~~~~~~~~~~~~~~~Original Magnitude} &
\multicolumn{5}{c}{~~~~~~~~~~~~~~~~~~~~~~~~~~~S-correction} & \multicolumn{1}{c}{S}  \\ 
 & (+2400000) & $U$ & $B$  & $V$ & $R$& $I$ & $\Delta U$& $\Delta B$& $\Delta V$& $\Delta R$& $\Delta I$& \\ \hline
 13/7 & 53564.50& 16.837 & 16.113  & 14.902  & 14.492  & 14.260   & -0.241& -0.018& 0.031& -0.004&  0.012 & 5 \\ 
 14/7 & 53566.41&        & 16.327  & 15.066  & 14.633  & 14.291   &       & -0.010&-0.024& -0.019&  0.035 & 7 \\ 
 18/7 & 53569.57& 16.831 & 16.592  & 15.179  & 14.863  & 14.444   & -0.065& -0.096& 0.056& -0.027&  0.044 & 8 \\ 
 20/7 & 53571.60& 16.945 & 16.684  & 15.298  & 14.977  & 14.599   & -0.066& -0.097& 0.054& -0.029&  0.044 & 8 \\ 
 20/7 & 53572.21&        & 16.616  & 15.430  & 15.071  & 14.650   & 	 & -0.023 &-0.009& -0.023&  0.072 & H \\
 22/7 & 53573.55& 16.965 & 16.753  & 15.384  & 15.097  & 14.727   & -0.066& -0.097& 0.051& -0.032&  0.045 & 8 \\ 
 23/7 & 53575.20& 16.753 & 16.660  & 15.546  & 15.204  & 14.810   &  0.190& -0.020&-0.008& -0.023&  0.076 & H \\
 24/7 & 53575.51& 17.002 & 16.799  & 15.471  & 15.193  & 14.827   & -0.066& -0.096& 0.048& -0.034&  0.046 & 8 \\ 
 27/7 & 53578.55& 17.086 & 16.891  & 15.574  & 15.327  & 14.977   & -0.065& -0.094& 0.043& -0.038&  0.048 & 8 \\ 
 30/7 & 53581.61& 17.152 & 16.914  & 15.659  & 15.417  & 15.129   & -0.064& -0.091& 0.039& -0.042&  0.051 & 8 \\ 
 31/7 & 53583.41& 16.996 & 16.862  & 15.755  & 15.443  & 15.299   &  0.182& -0.014& 0.016&  0.011& -0.018 & 3 \\ 
 01/8 & 53584.20&        & 16.833  & 15.823  & 15.543  & 15.275   &  	 & -0.022 &-0.010& -0.020&  0.084 & H \\
 04/8 & 53587.14& 16.997 & 16.897  & 15.881  & 15.660  & 15.378   &  0.201& -0.025&-0.012& -0.018&  0.086 & H \\
 05/8 & 53587.56& 17.258 & 17.005  & 15.873  & 15.619  & 15.392   & -0.062& -0.092& 0.033& -0.044&  0.056 & 8 \\ 
 08/8 & 53590.56& 17.300 & 17.076  & 15.929  & 15.733  & 15.522   & -0.061& -0.092& 0.031& -0.044&  0.059 & 8 \\ 
 08/8 & 53591.13&        & 16.970  & 16.006  & 15.763  & 15.575   &  	 & -0.027 &-0.015& -0.016&  0.089 & H \\  
 13/8 & 53595.50& 17.369 & 17.094  & 16.075  & 15.868  & 15.681   & -0.061& -0.089& 0.028& -0.043&  0.062 & 8 \\ 
 14/8 & 53597.10&        &         &         & 15.978  &          &  	 &        &      & -0.012&        & H \\
 15/8 & 53598.38& 17.275 & 17.096  & 16.172  & 15.943  & 15.886   &  0.187& -0.008& 0.014&  0.018&  0.009 & 3 \\ 
 17/8 & 53600.18&        &         &         &         & 15.992   &  	 &        &      &       &  0.096 & H \\
 21/8 & 53604.15&        & 17.192  & 16.368  & 16.173  & 16.109   &  	 & -0.022 &-0.020& -0.008&  0.099 & H \\  
 28/8 & 53611.38& 17.410 & 17.294  & 16.475  & 16.372  & 16.518   &  0.235& -0.003&-0.008& -0.001&  0.091 & 6 \\ 
 30/8 & 53613.37& 17.576 & 17.293  & 16.519  & 16.385  & 16.464   &  0.183&  0.023& 0.008&  0.020&  0.036 & 3 \\ 
 01/9 & 53615.14&        &         & 16.618  & 16.525  & 16.561   &  	 &        &-0.020&  0.001&  0.109 & H\\  
 03/9 & 53617.10&        & 17.341  & 16.674  & 16.552  & 16.596   &       &  0    &-0.020&  0.003&  0.112 & H\\
 07/9 & 53621.11& 17.741 & 17.412  & 16.798  & 16.695  & 16.754   &  0.183&  0.009&-0.020&  0.007&  0.116 & H\\
\hline
\end{longtable}
\begin{centering}
1 = 40inch SSO Telescope + WFI;\\
2 = 3.5m Telescopio Nazionale Galileo + DOLORES; \\
3 = 2.5m Nordic Optical Telescope + ALFOSC;\\
4 = 2.3m SSO Telescope + Imager;\\
5 = 1.2m Mercator Telescope + MEROPE;\\
6 = 2.2m Calar Alto Telescope + CAFOS;\\
7 = 1.82m Copernico Telescope + AFOSC;\\
8 = 2.2m ESO/MPI Telescope + WFI;  \\
H = 2m Himalayan Chandra Telescope + HFOSC.\\
\end{centering}

\newpage

\begin{longtable}{cccccccc}
\caption{Final, S--corrected $U$, $B$, $V$, $R$, $I$ magnitudes of SN
2005cf. The uncertainties reported in brackets
take into account both measurement and photometric calibration
errors. Unfiltered measurements from IAU Circ. 8534 are also reported.}\\
\hline\hline
\label{SN_mags_corr}
Date & JD & $U$ & $B$ & $V$ & $R$ & $I$ & S\\
& (+2400000) & & & & & &\\ \hline
 25/5 & 53515.87&   &   &  & $\geq$18.5 &   & 0 \\
 28/5 & 53518.86&   &   &  & 16.4 &   & 0 \\
 30/5 & 53520.85&   &   &  & 15.5 &   & 0 \\
31/5 & 53521.90 & & 15.462 (0.010) & 15.238 (0.008) & 15.138 (0.009) & 15.177  (0.011) & 1\\
31/5 & 53522.38 & 15.725 (0.043) & 15.272 (0.015) & 15.097 (0.010) & 14.976 (0.011) & 14.999  (0.024) & 2\\
01/6 & 53522.97 & & 15.012 (0.009) & 14.898 (0.008) & 14.780 (0.008) & 14.805  (0.009) & 1\\
01/6 & 53523.16 & 15.281 (0.055) & 14.924 (0.017) & 14.902 (0.020) & 14.771 (0.013) & 14.789  (0.024) & H\\
02/6 & 53524.13 & 14.847 (0.053) & 14.600 (0.024) & 14.649 (0.024) & 14.467 (0.016) & 14.501  (0.022) & H\\
02/6 & 53524.44 & 14.641 (0.044) & 14.561 (0.011) & 14.507 (0.009) & 14.383 (0.009) & 14.438  (0.010) & 3\\
03/6 & 53524.97 & & 14.392 (0.010) & 14.379 (0.007) & 14.236 (0.008) & 14.269 (0.009) & 1\\
03/6 & 53525.25 & & 14.339 (0.041) & 14.367 (0.015) & 14.224 (0.011) & 14.242 (0.020) & H\\
04/6 & 53526.35 & & 14.131 (0.041) & 14.185 (0.034) & 14.007 (0.012) & 14.035 (0.023) & H\\ 
06/6 & 53527.57 & 13.681 (0.045) & 13.899 (0.015) & 13.909 (0.013) & 13.822 (0.010) & 13.866  (0.023) & 2\\
06/6 & 53527.95 & & 13.862 (0.012) & 13.894 (0.010) & 13.730 (0.011) & 13.802 (0.013) & 1\\
06/6 & 53528.14 & & 13.841 (0.012) & 13.943 (0.021) & 13.788 (0.017) & 13.849 (0.018) & H\\
07/6 & 53529.24 & 13.596 (0.048) & 13.724 (0.011) & 13.824 (0.020) & 13.675 (0.011) & 13.785  (0.019) & H\\
08/6 & 53530.38 & 13.449 (0.044) & 13.694 (0.011) & 13.698 (0.008) & 13.556 (0.009) & 13.697  (0.009) & 3\\
08/6 & 53530.38 & 13.533 (0.062) & 13.670 (0.018) & 13.756 (0.021) & 13.610 (0.014) & 13.745  (0.030) & H\\
09/6 & 53531.12 & & 13.609 (0.029) & 13.699 (0.024) & 13.590 (0.014) & 13.713 (0.020) & H\\
09/6 & 53531.48 & & 13.642 (0.011) & 13.617 (0.011) & & 13.703 (0.010) & 3\\
10/6 & 53531.52 & & & & 13.514 (0.013) & & 3\\
10/6 & 53532.19 & & 13.597 (0.039) & 13.632 (0.020) & 13.545 (0.013) & 13.714 (0.027) & H\\
11/6 & 53533.13 & 13.487 (0.047) & 13.535 (0.029) & 13.622 (0.028) & 13.533 (0.019) & 13.720 (0.018) & H\\
12/6 & 53533.50 & 13.386 (0.044) & 13.587 (0.010) & 13.580 (0.009) & 13.447 (0.011) & 13.664 (0.010) & 3\\
12/6 & 53533.99 & 13.467 (0.029) & 13.539 (0.010) & 13.523 (0.011) & 13.455 (0.012) & 13.727 (0.047) & 4\\
14/6 & 53536.47 & 13.571 (0.065) & 13.594 (0.010) & 13.523 (0.008) & 13.445 (0.015) & 13.806 (0.035) & 5\\
16/6 & 53538.19 & 13.757 (0.056) & & 13.601 (0.022) & 13.551 (0.016) & 13.862 (0.019) & H\\
16/6 & 53538.36 & 13.695 (0.049) & 13.689 (0.013) & 13.558 (0.013) & 13.566 (0.015) & 13.846 (0.030) & 6\\
17/6 & 53539.21 & & 13.720 (0.022) & 13.633 (0.027) & 13.577 (0.016) & 13.897 (0.016) & H\\
17/6 & 53539.37 & 13.794 (0.049) & 13.764 (0.013) & 13.594 (0.008) & 13.571 (0.010) & 13.864 (0.028) & 6\\
18/6 & 53540.13 & & 13.767 (0.019) & 13.654 (0.026) & 13.620 (0.015) & 13.966 (0.023) & H\\
18/6 & 53540.41 & 13.892 (0.049) & 13.855 (0.013) & 13.635 (0.011) & 13.611 (0.012) & 13.952 (0.031) & 6\\
19/6 & 53541.37 & 13.979 (0.049) & 13.899 (0.016) & 13.693 (0.015) & 13.657 (0.012) & 14.008 (0.032) & 6\\
20/6 & 53542.26 & 14.067 (0.057) & 13.921 (0.024) & 13.727 (0.015) & 13.739 (0.032) & 14.071 (0.024) & H\\
21/6 & 53542.53 & 14.032 (0.050) & 14.115 (0.018) & 13.705 (0.015) & 13.700 (0.013) & 14.074 (0.030) & 6\\
21/6 & 53543.14 & 14.178 (0.060) & 14.045 (0.022) & 13.763 (0.025) & 13.813 (0.014) & 14.126 (0.018) & H\\
22/6 & 53543.51 & 14.304 (0.065) & 14.053 (0.009) & 13.727 (0.009) & 13.736 (0.015) & 14.179 (0.022) & 5\\
23/6 & 53545.13 & 14.423 (0.053) & 14.191 (0.025) & 13.892 (0.024) & 13.907 (0.016) & 14.259 (0.033) & H\\ 
24/6 & 53546.30 & 14.475 (0.052) & 14.323 (0.017) & 13.949 (0.022) & 14.033 (0.026) & 14.316 (0.030) & H\\
24/6 & 53546.36 & 14.513 (0.049) & 14.358 (0.016) & 13.909 (0.011) & 13.978 (0.015) & 14.265 (0.030) & 6\\
24/6 & 53546.43 & 14.583 (0.044) & 14.361 (0.011) & 13.940 (0.012) & 13.972 (0.010) & 14.307 (0.017) & 3\\
25/6 & 53546.50 & 14.682 (0.065) & 14.362 (0.009) & 13.915 (0.008) & 13.968 (0.015) & & 5\\
25/6 & 53547.45 & 14.628 (0.049) & 14.495 (0.012) & 13.986 (0.009) & 14.064 (0.010) & 14.310 (0.029) & 6\\
26/6 & 53548.40 & 14.731 (0.050) & 14.597 (0.014) & 13.981 (0.009) & 14.110 (0.012) & 14.323 (0.031) & 6\\
28/6 & 53550.37 & 15.076 (0.049) & 14.844 (0.013) & 14.159 (0.009) & 14.175 (0.008) & 14.332 (0.028) & 6\\
29/6 & 53551.48 & 15.398 (0.065) & 14.900 (0.009) & 14.205 (0.007) & 14.139 (0.015) & 14.403 (0.022) & 5\\
01/7 & 53553.37 & 15.493 (0.050) & 15.184 (0.013) & 14.334 (0.013) & 14.246 (0.015) & 14.306 (0.032) & 6\\
02/7 & 53554.37 & 15.601 (0.050) & 15.265 (0.014) & 14.359 (0.013) & 14.226 (0.016) & 14.283 (0.032) & 6\\
03/7 & 53555.47 & 15.890 (0.065) & 15.340 (0.009) & 14.419 (0.008) & 14.201 (0.015) & 14.207 (0.021) & 5\\
06/7 & 53558.48 & 16.111 (0.066) & 15.616 (0.009) & 14.590 (0.008) & 14.291 (0.015) & 14.197 (0.022) & 5\\
07/7 & 53559.19 & 16.140 (0.050) & 15.726 (0.018) & 14.660 (0.023) & 14.320 (0.011) & 14.192 (0.015) & H\\
07/7 & 53559.49 & 16.225 (0.044) & 15.797 (0.010) & 14.614 (0.008) & 14.287 (0.010) & 14.188 (0.009) & 3\\ 
10/7 & 53562.48 & 16.411 (0.065) & 15.952 (0.010) & 14.794 (0.008) & 14.401 (0.015) & 14.147 (0.022) & 5\\
11/7 & 53563.24 & 16.404 (0.059) & 16.053 (0.013) & 14.874 (0.026) & 14.477 (0.018) & 14.157 (0.029) & H\\ 
		\\	    	        	     		    
\caption{continued.}\\
\hline\hline
Date & JD & $ U$ & $ B$ & $ V$ & $ R$ & $ I$ & S\\
& (+2400000) & & & & & &\\ \hline
11/7 & 53563.37 & 16.469 (0.051) & 16.069 (0.013) & 14.861 (0.010) & 14.469 (0.011) & 14.255 (0.031) & 6\\
12/7 & 53564.37 & 16.539 (0.049) & 16.158 (0.013) & 14.934 (0.010) & 14.525 (0.009) & 14.294 (0.029) & 6\\
13/7 & 53564.50 & 16.596 (0.066) & 16.095 (0.010) & 14.933 (0.008) & 14.488 (0.015) & 14.272 (0.021) & 5\\
14/7 & 53566.41 & & 16.317 (0.015) & 15.042 (0.017) & 14.614 (0.021) & 14.325 (0.026) & 7\\
18/7 & 53569.57 & 16.766 (0.032) & 16.496 (0.012) & 15.235 (0.009) & 14.836 (0.009) & 14.488 (0.018) & 8\\
20/7 & 53571.60 & 16.879 (0.030) & 16.587 (0.011) & 15.351 (0.009) & 14.948 (0.009) & 14.643 (0.018) & 8\\
20/7 & 53572.21 & & 16.593 (0.022) & 15.421 (0.031) & 15.048 (0.019) & 14.721 (0.025) & H\\
22/7 & 53573.55 & 16.899 (0.030) & 16.656 (0.011) & 15.435 (0.009) & 15.065 (0.008) & 14.772 (0.017) & 8\\
23/7 & 53575.20 & 16.943 (0.063) & 16.640 (0.021) & 15.538 (0.028) & 15.181 (0.021) & 14.885 (0.023) & H\\
24/7 & 53575.51 & 16.937 (0.029) & 16.703 (0.011) & 15.519 (0.009) & 15.159 (0.009) & 14.873 (0.018) & 8\\
27/7 & 53578.55 & 17.021 (0.030) & 16.797 (0.011) & 15.617 (0.009) & 15.289 (0.009) & 15.025 (0.017) & 8\\
30/7 & 53581.61 & 17.088 (0.030) & 16.823 (0.012) & 15.698 (0.009) & 15.376 (0.009) & 15.180 (0.018) & 8\\
31/7 & 53583.41 & 17.178 (0.044) & 16.849 (0.010) & 15.771 (0.008) & 15.454 (0.009) & 15.281 (0.009) & 3\\
01/8 & 53584.20 & & 16.811 (0.027) & 15.813 (0.028) & 15.523 (0.011) & 15.359 (0.035) & H\\
04/8 & 53587.14 & 17.198 (0.052) & 16.872 (0.020) & 15.869 (0.020) & 15.642 (0.027) & 15.464 (0.018) & H\\
05/8 & 53587.56 & 17.196 (0.030) & 16.913 (0.011) & 15.906 (0.009) & 15.575 (0.009) & 15.448 (0.018) & 8\\
08/8 & 53590.56 & 17.239 (0.029) & 16.984 (0.011) & 15.960 (0.009) & 15.689 (0.009) & 15.580 (0.018) & 8\\
08/8 & 53591.13 & & 16.943 (0.013) & 15.991 (0.022) & 15.747 (0.013) & 15.664 (0.026) & H\\
13/8 & 53595.50 & 17.308 (0.109) & 17.005 (0.022) & 16.103 (0.040) & 15.825 (0.020) & 15.743 (0.027) & 8\\
14/8 & 53597.10 & & & & 15.966 (0.018) & & H\\
15/8 & 53598.38 & 17.462 (0.045) & 17.088 (0.010) & 16.186 (0.009) & 15.961 (0.010) & 15.895 (0.009) & 3\\
17/8 & 53600.18 & & & & & 16.088 (0.021) & H\\
21/8 & 53604.15 & & 17.171 (0.019) & 16.348 (0.031) & 16.165 (0.011) & 16.208 (0.022) & H\\ 
28/8 & 53611.38 & 17.645 (0.051) & 17.291 (0.013) & 16.467 (0.011) & 16.371 (0.009) & 16.609 (0.030) & 6\\
30/8 & 53613.37 & 17.759 (0.045) & 17.316 (0.012) & 16.527 (0.011) & 16.405 (0.011) & 16.500 (0.011) & 3\\
01/9 & 53615.14 & & & 16.598 (0.019) & 16.526 (0.014) & 16.670 (0.030) & H\\
03/9 & 53617.10 & & 17.341 (0.018) & 16.654 (0.028) & 16.555 (0.015) & 16.707 (0.018) & H\\
07/9 & 53621.11 & 17.924 (0.053) & 17.421 (0.021) & 16.778 (0.025) & 16.702 (0.015) & 16.870 (0.025) & H\\ \hline
\end{longtable}
\begin{centering}
0 = unfiltered magnitudes from IAU Circ. 8534 \\
1 = 40inch SSO Telescope + WFI;\\
2 = 3.5m Telescopio Nazionale Galileo + DOLORES; \\
3 = 2.5m Nordic Optical Telescope + ALFOSC;\\
4 = 2.3m SSO Telescope + Imager;\\
5 = 1.2m Mercator Telescope + MEROPE;\\
6 = 2.2m Calar Alto Telescope + CAFOS;\\
7 = 1.82m Copernico Telescope + AFOSC;\\
8 = 2.2m ESO/MPI Telescope + WFI;  \\
H = 2m Himalayan Chandra Telescope + HFOSC.\\
\end{centering}
\end{center}
\normalsize
\twocolumn
\newpage

We remark that S--correction in the $U$ band is affected by a
non--negligible uncertainty due to the low quantum
efficiency of the CCDs  and errors in the flux calibration
of the SN spectra below $\sim$3500 \AA.

The comparison between the $U$ and the $I$ band light curves of
SN~2005cf (Fig. \ref{corr_uncorr}, top and bottom, respectively) before and 
after the S--correction, displays the improvement in the quality of the photometry. 
The final optical light curves are shown in Fig. \ref{fig:lightcurves}, while the 
S-corrected magnitudes of SN 2005cf are reported in Tab. \ref{SN_mags_corr}.
 
Hereafter, we will refer to JD = 2453534.0 as the epoch of the $B$ band
maximum light (see Sect. \ref{sect-param}).

\begin{figure}
\resizebox{\hsize}{!}{\includegraphics{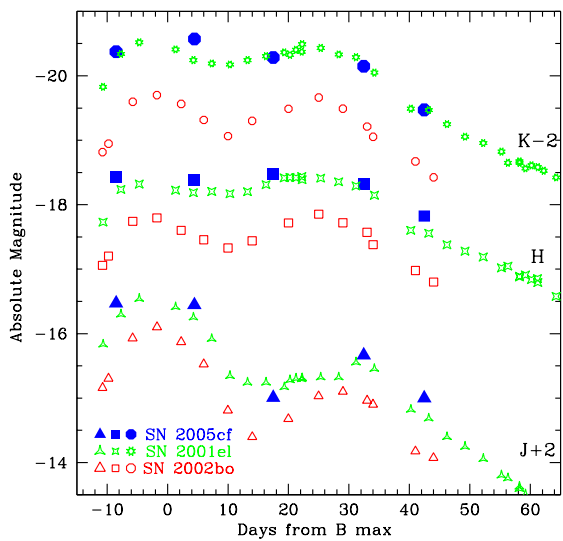}}
\caption{NIR absolute light curves of SNe 2005cf (blue filled symbols),
2001el (green starred symbols) and 2002bo (red open symbols). The sources of
the data are cited in the text.  $K'$ photometry is displayed for SN~2005cf.}
\label{fig:IRlc}
\end{figure}

\subsection{Near--IR Light Curves}

Contrary to the optical photometry, no S--correction was applied 
to our NIR photometry, owing to the lack of adequate time coverage
of the NIR spectroscopy. The NIR photometry available for SN~2005cf 
is shown in Tab. \ref{SNIR_mags}. 
In Fig. \ref{fig:IRlc}, the absolute NIR light curves of SN~2005cf are compared
with those of the well--studied SNe 2001el \citep{kris03} and 2002bo \citep{kris04}.
The absolute magnitudes were computed assuming the distance
modulus and total reddening values of Tab. \ref{SN_cfr}
(see Sect. \ref{color_and_bolo}).

\begin{table}
\caption{$J$, $H$, $K'$ magnitudes of SN 2005cf and assigned errors.}
\centering
\label{SNIR_mags}
\begin{tabular}{cccccc}
\hline\hline
Date & JD & $J$ & $H$ & $K'$ & S\\ 
& (+2400000) & & & &\\ \hline
03/6 & 53525.49 &  14.13 (0.02) & 14.13 (0.03) & 14.17 (0.03) & A\\
16/6 & 53538.45 &  14.15 (0.03) & 14.18 (0.03) & 13.97 (0.03) & A\\
29/6 & 53551.48 &  15.59 (0.02) & 14.10 (0.03) & 14.26 (0.03) & B\\
14/7 & 53566.45 &  14.93 (0.02) & 14.25 (0.03) & 14.40 (0.03) & A\\ 
24/7 & 53576.47 &  15.60 (0.03) & 14.75 (0.03) & 15.07 (0.03) & A\\ 
\hline
\end{tabular}

A = Telescopio Nazionale Galileo 3.5m + NICS; \\
B = Calar Alto 3.5m Telescope + Omega--Cass\\
\end{table}

SN~2002bo is significantly fainter than both SN~2005cf and
2001el. However, we remark that the behaviour of SN~2002bo in the NIR 
is rather peculiar and that SN~2005cf was observed in the $K'$ band, while SN~2001el 
and SN~2002bo were calibrated by \citet{kris03,kris04} in the standard
Persson's system \citep{pers98}. The deep minimum in the $J$ band light
curve of SN~2005cf resembles that observed in SN~2002bo, although the
$J$ band luminosity is closer to that of SN~2001el. The
plateau--like behaviour of the $H$ band light curve of SN~2005cf between phase about
$-$10 and +30 is very similar to that observed in the light curve of SN~2001el.
A strong similarity between SN~2005cf and SN~2001el is also seen in the 
$K$ band evolution. Their $K$ band light curves remain relatively flat
from phase $-$10 and +30, while the maxima of SN~2002bo are somewhat 
more pronounced. 

Recently, \citet{kas06} explained the variable strength
of the NIR secondary maximum in SNe Ia in terms of different 
abundance stratification, metallicities of the progenitor star and
amounts iron--group elements synthesized in the explosion.
In particular, Type Ia SNe ejecting more radioactive $^{56}$Ni are expected to 
show, together with a brighter light curve, more pronounced NIR secondary maxima.
 
\subsection{Reddening and Distance}   \label{red}

SN 2005cf exploded very far from the nucleus of MCG --01--39--003,
in a region with low background contamination. 
Deep late--time VLT imaging (F. Patat, private
communication) shows that the SN exploded at the
edge of a long tidal bridge connecting MCG --01--39--003
with the interacting companion \citep[see also][]{voro63}, 
suggesting relatively
small host galaxy interstellar extinction. This finding is supported by
non--detection of narrow interstellar lines
in the SN spectra. As Sect. \ref{color_and_bolo} will show, a small value for the
interstellar extinction is also supported by the normal colour curves of SN~2005cf.
 Therefore, in this paper we
will adopt as total extinction the Galactic estimate at the coordinates
of the SN, i.e. E$(B-V)$ = 0.097$\pm$0.010, reported by \citet{schl98}.

Despite the relatively short distance, the galaxy system hosting
SN~2005cf is poorly
studied. As a consequence, large uncertainty exists in the distance
estimate.
For MCG --01--39--003 and NGC 5917 LEDA provides the recession velocities
corrected for the effects of the  Local Group infall onto 
the Virgo Cluster \citep[208 km s$^{-1}$,][]{terry02}: v$_{Vir}$ =  
1977 and 1944 km s$^{-1}$ are reported for the two galaxies, respectively.
However, we should take into account a non--negligible gravitational
effect of the Virgo Cluster at the distance of the two galaxies. 
Taking into account the observed positions of the two galaxies relative
to the Virgo centre, a crude estimate of the virgocentric
component subtracts to the observed recession velocities about
3--400 km s$^{-1}$. 
hereafter we will adopt the first infall velocity model.

\begin{figure*}   
\resizebox{\hsize}{!}{
 \includegraphics{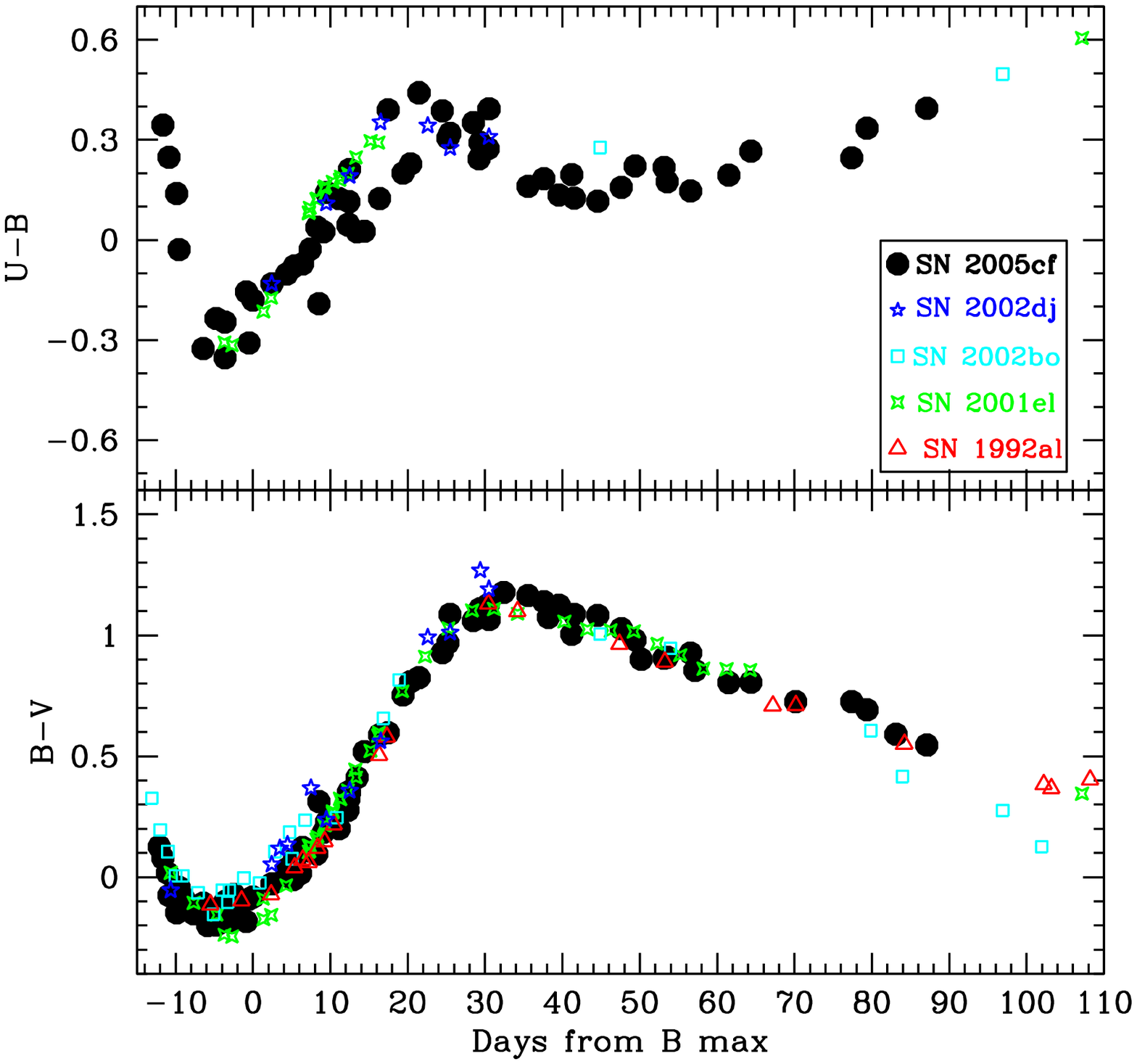}
 \includegraphics{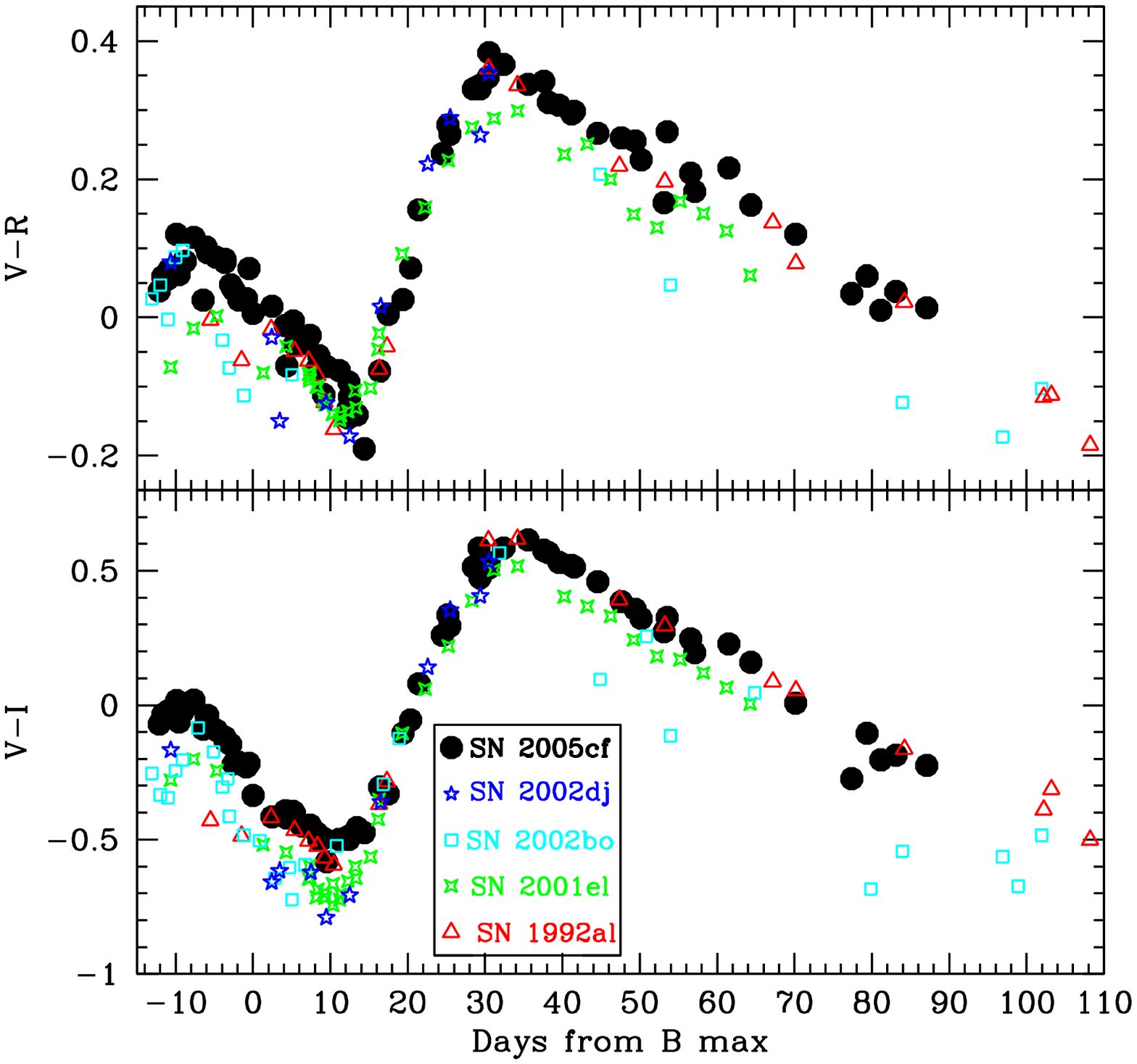}}
   \caption{Colour evolution of SN~2005cf compared with other SNe Ia
with similar $\Delta m_{15}(B)_{true}$: SNe 1992al, 2001el, 
2002bo, 2002dj. Evolution of $U-B$ (top--left), $B-V$
 (bottom--left), $V-R$ (top--right) and $V-I$ (bottom--right) colours.
All colour curves are reddening corrected.
For references, see the text.}
 \label{fig:color}
\end{figure*}

\begin{table*}
\caption{Basic information for the SNe Ia sample with $\Delta
m_{15}(B)_{true}$ $\sim$1.1 included in this paper.}
\centering
\label{SN_cfr}
\begin{tabular}{cccccccc}
\hline\hline
 SN  & host galaxy & $\mu$ & E$(B-V)_{TOT}^{\otimes}$ & JD$(B)_{max}$ & $M_{B,max}$ & $\Delta m_{15}(B)_{true}$ & sources \\ \hline
2005cf &MCG --01--39--003 & 32.51 & 0.097 & 2453534.0  & -19.39 & 1.12 & 1,0 \\
2002dj &NGC 5018 & 32.92 & 0.15  & 2452450.5  & -19.17 & 1.15 & 2,0 \\
2002bo &NGC 3190 & 31.45 & 0.38  & 2452356.5  & -18.98  & 1.17 & 3,4,0 \\
2001el &NGC 1448 & 31.29 & 0.22$^{\ddag}$ & 2452182.5 & -19.35 & 1.13 & 5  \\
1992al &ESO 234--G069 & 33.82 & 0.034 & 2448838.36 & -19.37 & 1.11 & 6,0 \\ \hline
\end{tabular}

$^{\otimes}$ E(B-V)$_{Gal}$ + E(B-V)$_{host}$;
$^{\ddag}$ E(B-V)$_{host}$ = 0.18, with R$_V$ = 2.88.\\
 0 = LEDA; 1 = this paper; 2 = Pignata et al., in preparation; 3 = \protect\cite{ben04}; \\
4 = \protect\cite{steh05}; 5 = \protect\cite{kris03}; 6 = \protect\cite{hamu96}.\\ 
\end{table*}

\citet{kran86} computed distance estimates of a large sample of nearby
galaxies based on a virgocentric non--linear flow model \citep[see
e.g.][]{silk77}. While  MCG --01--39--003 and NGC 5917 are not listed in the catalogue,
with a good approximation we can assume the same infall 
correction computed for another galaxy, NGC 5812, which projects very close to the SN~2005cf
parent galaxy and has similar recession velocity (v$_{Vir}$ = 1965 km $s^{-1}$,
LEDA, cf. Tab. \ref{gal_param}).
In the Kraan--Korteveg's catalogue, the distances are expressed
in units of the Virgo Cluster distance (d$_{Vir}$). For NGC 5812
two alternative distances are derived from different assumptions on
the virgocentric infall velocity of the Local Group: 1.91 d$_{Vir}$
for a more commonly accepted 220 km s$^{-1}$ -- model \citep{tamm85}
or 1.84 d$_{Vir}$ for a 440  km s$^{-1}$ -- model \citep{kran86}.
Since a value for the local infall velocity toward Virgo of 220  km s$^{-1}$
is close to that currently adopted by LEDA (208 km s$^{-1}$),

To derive the distance modulus of SN~2005cf we need to adopt a
distance for Virgo. For the latter, the values reported in the literature show
some scatter. For instance, the mean Tully--Fisher distance of Virgo 
obtained by \citet{fou01} from 51 spiral galaxies members of the cluster
is  d = 18.0 $\pm$ 1.2 Mpc. This gives a distance of NGC 5812 of 34.4
Mpc ($\mu$ = 32.68).

Alternatively, computing the cepheid distances of 6 galaxies of the Virgo Cluster,
 \citet{fou01} found a somewhat smaller distance of Virgo: d = 15.4 $\pm$ 0.5 Mpc.
The resulting distance of NGC 5812 is 29.4 Mpc ($\mu$ = 32.34). 
Averaging the  distance moduli of NGC 5812 obtained from the two different
estimates of the Virgo distance, we obtain $\mu$ = 32.51. We can
reasonably adopt this distance modulus also for MCG --01--39--003 and
NGC 5917. 
A conservative estimate of the error is
obtained from the dispersion of the galaxy peculiar motions, i.e.
$\sim$350 km s$^{-1}$ \citep{som97}, which gives a maximum error
in the distance modulus of $\Delta$$\mu$ = 0.33.
Hereafter, we will adopt $\mu$ = 32.51 $\pm$ 0.33 as our 
best distance modulus estimate for the galaxy hosting SN~2005cf.

\subsection{Colour Curves, Absolute Luminosity and Bolometric Light Curve} 
\label{color_and_bolo}

In what follows, we compare colour evolution, absolute light curves
and pseudo--bolometric luminosity of SN~2005cf with those of other Type Ia
SNe with similar light curve shape.
The range of $\Delta m_{15}(B)$ $\sim$ 1.0--1.2 (around the average value
for normal SNe Ia) is well populated. As comparison 
objects  we have selected SNe 1992al \citep{hamu96}, 2001el \citep{kris03}, 
2002bo \citep{ben04,kris04} and 2002dj (Pignata et al., in preparation). 
Basic information about the distance moduli and reddening values
adopted for this sample is reported in Tab. \ref{SN_cfr}. In particular,
as already seen in Sect. \ref{red}, $\mu$ = 32.51
and E$(B-V)$ = 0.097 \citep{schl98} were adopted for SN~2005cf. 

In Fig. \ref{fig:color} the $U-B$ (top--left), $B-V$ (bottom--left),
$V-R$ (top--right) and $V-I$ (bottom--right) intrinsic colour curves of SN~2005cf and the other
similar SNe Ia mentioned above are shown.
The colour evolution of all these objects is very similar, with a few
minor differences. The $U-B$ colour (Fig. \ref{fig:color},
top--left) has a steep decline (from 0.4 to
about $-$0.3) until $\sim$4 days before maximum. Then the $U-B$ colour
curves become redder, arriving at $\sim$0.3 about 3--4 weeks past maximum
and being almost constant thereafter. SNe~2001el and 2002dj have 
a similar evolution.

The evolution of the $B-V$ colour curves (Fig. \ref{fig:color}, bottom--left)
is very similar for all SNe of our sample, showing a decreasing trend
from $\sim$ 0.3 to $-$0.2 in the period 13 to 5 days before the $B$ band
maximum. Then the $B-V$ colour rises for approximately one month, reaching
a $B-V$ $\approx$ 1.2. In the subsequent two months, the $B-V$ colour becomes
bluer again, to values below $\sim$0.5 at a phase of $\sim90$ days past maximum.

The $V-R$ and $V-I$ colours show a similar behaviour (see Fig. \ref{fig:color},
top--right and bottom--right). Soon after the explosion, the colours 
turn red. Then, between about a week before and two weeks past
maximum, the trend is reversed (the $V-I$ colour, in particular,
decreases from 0 to $-$0.7 in this time interval).
Subsequently, until about 1 month past the $B$ band maximum, the colours
become again redder ($V-R$ rises from $-$0.2 to +0.4, $V-I$ from $-$0.7
to 0.6), followed by a phase where they turn bluer again. 
About 3 months after maximum, the V--R colour reaches 0 and V--I about $-$0.3, 
with an ongoing trend to bluer values in the subsequent weeks.
The only outlier is SN 2002bo, which seems to have bluer $V-R$ and $V-I$
colours at all phases, especially well after maximum.

\begin{figure}   
\resizebox{\hsize}{!}{
 \includegraphics{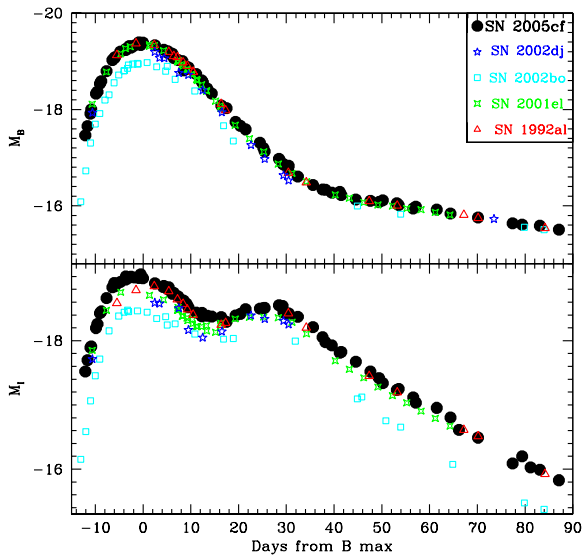}}
   \caption{$B$ (top) and $I$ band (bottom) absolute light curves for SN~2005cf and other
 similar objects: SNe 1992al, 2001el, 
2002bo, 2002dj. The values of $\mu$ and E$(B-V)$ adopted are
 shown in Tab. \ref{SN_cfr}.}
 \label{fig:absolute}
\end{figure}

A comparison of the absolute light curves of SN 2005cf and other similar 
events is shown in Fig. \ref{fig:absolute}, both for the $B$ band 
(top panel) and the $I$ band (bottom panel).  It is remarkable that the
$I$ band secondary maximum, known to be more or less pronounced
in different SNe Ia, is similarly prominent for the SNe of this sample.
The $B$ band maximum magnitude of SN~2005cf 
is $M_{B,max}$ $\approx -$19.39, which is 
in the bright side of the SN Ia luminosity function. 
At the epoch of the $B$ band maximum, the 
dereddened colour is $(B-V)_{B,max} = -$0.09.

\begin{figure}   
\resizebox{\hsize}{!}{
 \includegraphics{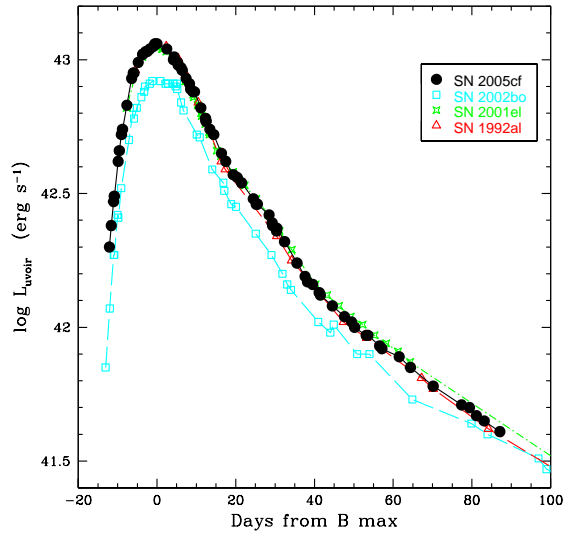}}
   \caption{Quasi--bolometric light curves for
 SNe~2005cf,  1992al, 2001el and 2002bo. 
All these SNe Ia are characterised by similar values for the $\Delta m_{15}(B)$
 parameter (about 1.1). }
 \label{fig:bolometric}
\end{figure}

The luminosity evolution of SN~2005cf obtained integrating the fluxes
in the optical bands is shown in Fig. \ref{fig:bolometric}. 
For comparison, also the observed pseudo--bolometric
light curves for other similar SNe Ia are shown.
Since $U$ band observations of SN~1992al are missing, we applied a $U$
band correction to its light curve following \citet{con00}. 
The pseudo--bolometric light curves of our Type Ia SN sample
are extremely similar, with the exception of SN~2002bo, 
which is fainter than SN~2005cf by a factor 0.75,
suggesting a non--negligible scatter in luminosity also among similarly
declining SNe Ia.

\section{SN Parameters} \label{sect-param}

The excellent photometric coverage of SN~2005cf allows us
to precisely estimate the epoch, and
the apparent and absolute magnitudes at the $B$, $V$ and $I$
maxima. The parameters for all bands are obtained by fitting the
light curves with a low--degree spline function.
The results are reported in Tab. \ref{SN_par1}. In particular,
the epoch of the $B$ band maximum is found to be JD($B_{max}$) =
2453534.0$\pm$0.3 (June 12.5 UT).

\begin{table}
\caption{Parameters of SN~2005cf derived from the optical light
curves. The errors in the absolute magnitudes are largely dominated 
by the uncertainty in the distance modulus estimate ($\pm$0.33). The values for
A$_{\lambda}$ are those provided by \protect\citet{schl98}.}
\centering
\label{SN_par1}
\begin{tabular}{cccccc}
\hline\hline
   & JD(max) & $m_{\lambda,max}$ & A$_{\lambda}$ & $M_{\lambda,max}$  & $\Delta m_{15}(\lambda)_{obs}$\\ 
   & (+24000000) & & & & \\ \hline
$ U$ & 53532.4$\pm$0.6 & 13.40$\pm$0.05 & 0.53 & $-$19.64 & 1.26 $\pm$ 0.05\\
$ B$ & 53534.0$\pm$0.3 & 13.54$\pm$0.02 & 0.42 & $-$19.39 & 1.11 $\pm$ 0.03\\
$ V$ & 53535.3$\pm$0.3 & 13.53$\pm$0.02 & 0.32 & $-$19.30 & 0.61 $\pm$ 0.02\\
$ R$ & 53534.6$\pm$0.4 & 13.45$\pm$0.03 & 0.26 & $-$19.32 & 0.71 $\pm$ 0.04\\  
$ I$ & 53532.0$\pm$0.5 & 13.70$\pm$0.03 & 0.19 & $-$19.00 & 0.61 $\pm$ 0.06\\
\hline
\end{tabular}
\end{table}
 
\begin{figure}   
\centering
{\includegraphics[width=6.0cm,angle=270]{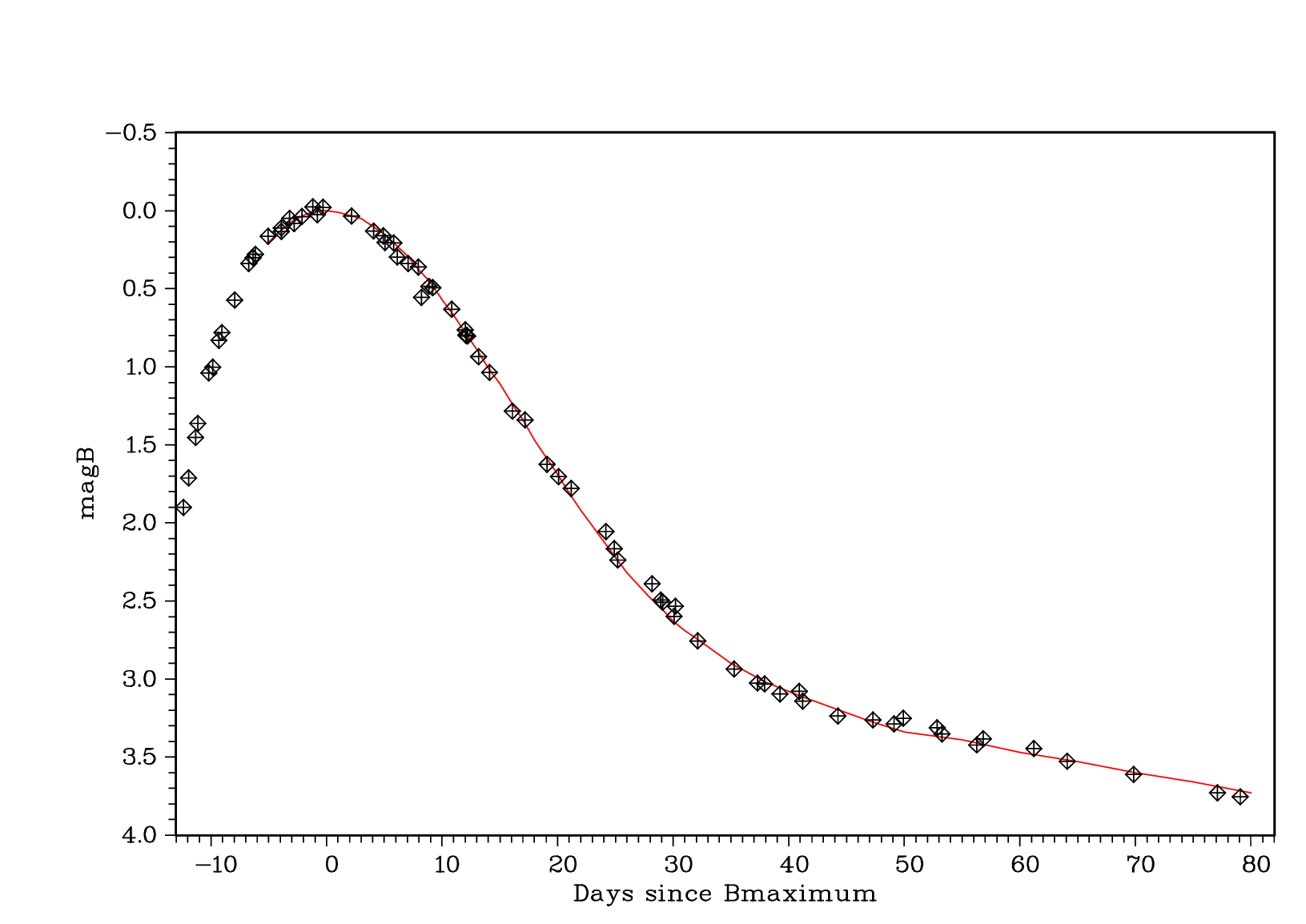}
 \includegraphics[width=6.0cm,angle=270]{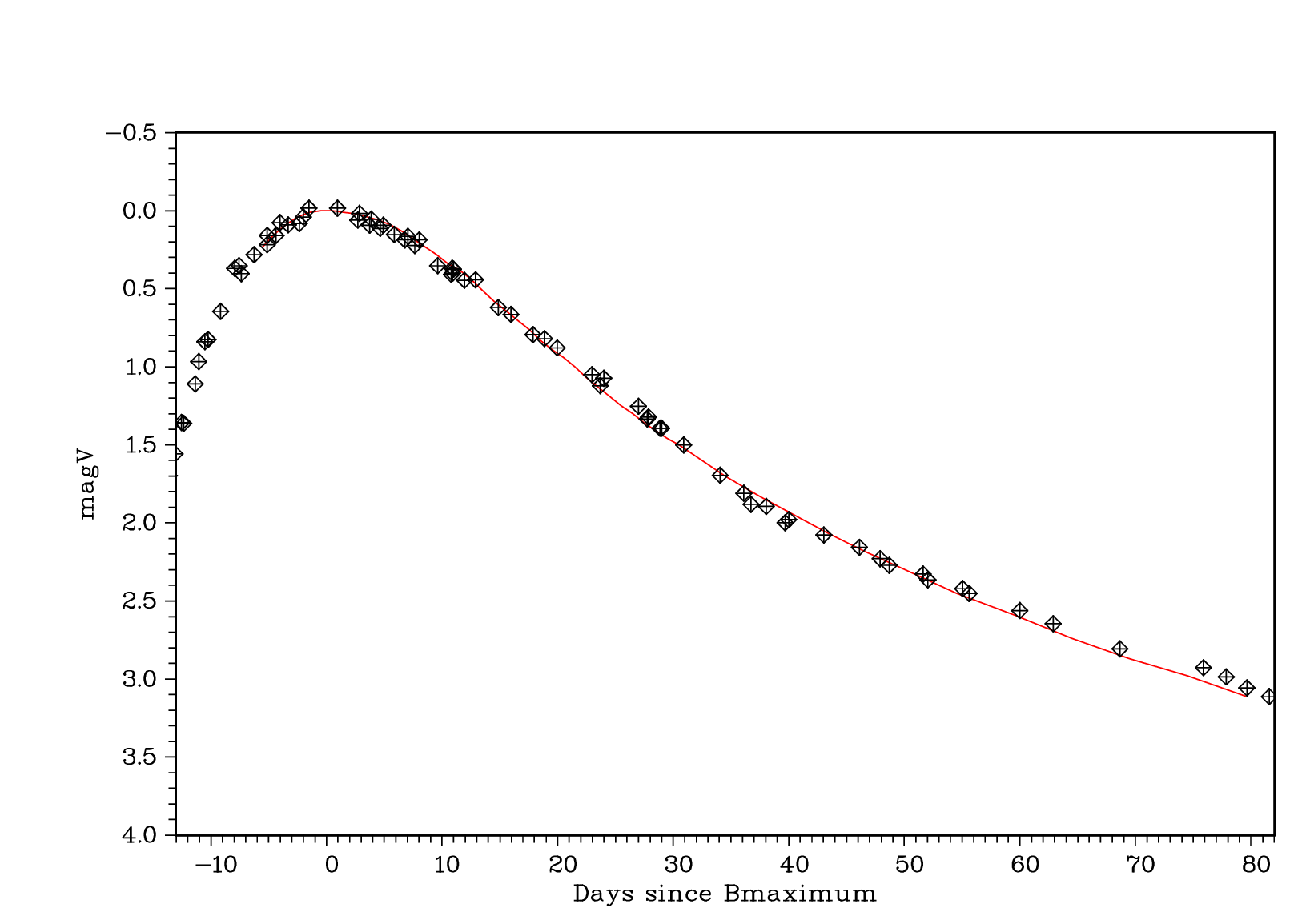}
 \includegraphics[width=6.0cm,angle=270]{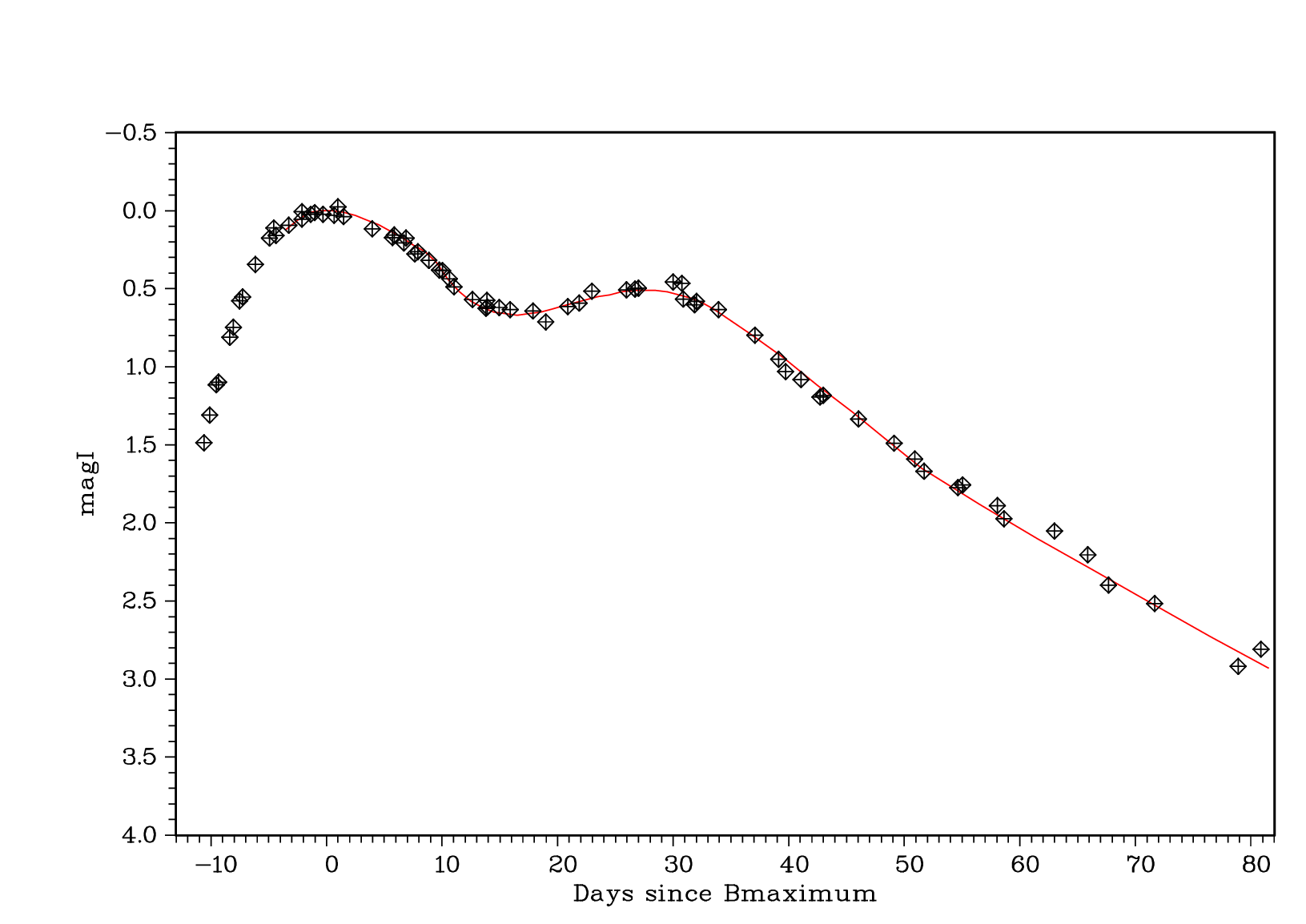}}
   \caption{Comparison between light curves of SN~2005cf
 (diamonds) and the template SN~1992al (solid line, from Hamuy et
 al., 1996), in the $B$, $V$, $I$
 bands (from top to bottom).}
 \label{sn05cf_vs_92al}
\end{figure}

In Fig. \ref{sn05cf_vs_92al} the B, V and I light curves 
of SN~2005cf are compared with those of the template SN~1992al \citep{hamu96}.  
The match of the light curves is excellent and therefore 
one of the most important parameters for SNe Ia, the $\Delta
m_{15}(B)$ is expected to be very similar. 
An observed $\Delta m_{15}(B)_{obs}$ $\approx$ 1.11 $\pm$ 0.03 is measured
for SN~2005cf (see column 6 in Tab. \ref{SN_par1}), well matching that derived for SN~1992al \citep{hamu96}.
This value makes SN~2005cf a typical SN Ia.
Owing to the low interstellar extinction suffered by SN~2005cf, the 
correction for reddening to apply to $\Delta m_{15}(B)$ is
very small. The reddening corrected $\Delta m_{15}(B)_{true}$ is obtained
applying the relation of \citet{phil99}:

\begin{equation}
\Delta m_{15}(B)_{true} = \Delta m_{15}(B)_{obs} + 0.1 \times E(B-V) 
\end{equation}
This gives $\Delta m_{15}(B)_{true}$ = 1.12. \\

An alternative parameter characterising the light curves of Type Ia SNe
is the stretch factor {\it s$^{-1}$} \citep{perl97}, i.e. the coefficient
indicating the stretch in time of the $B$ band light curve. We compute
for SN~2005cf $s^{-1}$ = 0.99$\pm$0.02. This result is in excellent agreement
with the value (0.995$\pm$0.179) derived applying the relation of \cite{alt04}:
\begin{equation}
\Delta m_{15}(B)_{true} = 1.98 \times (s^{-1} - 1) + 1.13
\end{equation} 

The $\Delta m_{15}(B)$--calibrated absolute magnitude at the
$B$ band maximum can be computed applying various relations
available in literature.
The relation between $M_{B,max}$ and $\Delta m_{15}(B)$ from \cite{hamu96}:

\begin{equation} \label{eq1}
M_{\lambda,max}^\star = A + B \times [\Delta m_{15}(B) - 1.1]
\end{equation}

provides a preliminary tool to get a calibrated absolute magnitude at maximum for SN~2005cf.  
The label $M_{\lambda,max}^\star$ is to indicate that this magnitude
has to be rescaled to  H$_0$ = 72 km s$^{-1}$ Mpc$^{-1}$.
Using the ``low extinction case'' parameters reported in Tab. 3 of
\cite{hamu96} and after rescaling $M_{B,max}^\star$ to H$_0$ = 72 km s$^{-1}$
Mpc$^{-1}$, we obtain $M_{B,max}$ =$-$19.03 $\pm$ 0.05. 
An updated, reddening--free decline rate vs. $\Delta m_{15}(B)$ relation was provided
by \cite{phil99} (see also their Tab. 3):

\begin{equation} \label{eq2}
M_{\lambda,max}^\star = M_{\lambda,max}^{1.1} + a \times [\Delta
m_{15}(B) - 1.1] + b \times [\Delta m_{15}(B) - 1.1]^2
\end{equation} 

From Eq. \ref{eq2},
using as $M_{\lambda,max}^{1.1}$ the coefficient A of Eq. \ref{eq1} and
reporting the magnitudes to H$_0$ = 72 km s$^{-1}$ Mpc$^{-1}$,
we derive for SN~2005cf a reddening--corrected absolute magnitude 
of $M_{B,max}$ = 19.02$\pm$0.05. The magnitudes for the other bands can 
be found in Tab. \ref{abs_max}.

\cite{alt04} used a further, updated version of Eq. \ref{eq1}, with different
coefficients ($A$ = $-$19.403 $\pm$ 0.044 and $B$ = 1.061 $\pm$ 0.154, under the
assumption of intermediate Cepheids metallicity, $\Delta$Y/$\Delta$Z = 2.5,
and with R$_B$ = 3.5). Applying the relation of Altavilla et al., we
obtain $M_{B,max}$ = $-$19.35 $\pm$ 0.06 (for H$_0$ = 72 km s$^{-1}$ Mpc$^{-1}$).

Using a large sample of SNe Ia, \cite{pri06} provided an updated
version of the relation between absolute $B$ band magnitude at peak
 and post--maximum decline (Eq. \ref{eq1}, but with different 
values for the coefficients A and B). Using the
coefficients of the low host galaxy extinction case (see \cite{pri06}, 
their Tab. 3, middle), we obtain for SN~2005cf $M_{B,max}$ = $-$19.31
$\pm$ 0.03. Estimates for other bands are reported in Tab. \ref{abs_max}.
The discrepancy of these magnitudes with those derived from 
other methods (see Tab. \ref{abs_max}) may be due to the uncertainty 
in the zeropoints of the Eq. \ref{eq1} reported in \cite{pri06}. 
Using different sub--samples, the scatter in the zeropoint values
is between 0.10 and 0.15 magnitudes in all bands  (see their Tab. 4). 

Another approach for estimating the absolute magnitude at maximum is
that of \cite{rein05}, who computed the value from $\Delta m_{15}(B)$
and colour at maximum. Rewriting their Eq. (23),

\begin{equation} \label{eq3}
M_{\lambda,max}^\star = \alpha \times [\Delta m_{15}(B) - 1.1] + \beta \times [(B-V)_0 + 0.024] + \gamma
\end{equation}

with the coefficients $\alpha$, $\beta$ and $\gamma$ reported in
Tab. 5 of \cite{rein05}, and with $(B-V)_0$ obtained from their
empirical relation

\begin{equation} \label{eq4}
(B-V)_0 = 0.045 \times \Delta m_{15}(B) - 0.073
\end{equation}

we obtain the absolute magnitudes shown in
Tab. \ref{abs_max}. For the $B$ band it is $-$19.16 $\pm$ 0.06
(with H$_0$ = 72 km s$^{-1}$ Mpc$^{-1}$).

\begin{table}
\caption{$\Delta m_{15}(B)$--corrected absolute magnitudes for
SN~2005cf. All absolute magnitudes are scaled to H$_0$ = 72 km s$^{-1}$ Mpc$^{-1}$.}
\centering
\label{abs_max}
\begin{tabular}{cccc}
\hline\hline
 method  & $M_{B,max}$ &$M_{V,max}$ &$M_{I,max}$ \\ \hline
Phillips et al. (1999)$^\ddag$ & $-$19.02$\pm$0.05 & $-$19.03$\pm$0.05 & $-$18.76$\pm$0.05 \\
Altavilla et al. (2004) & $-$19.35$\pm$0.06 & & \\
Prieto et al. (2006) & $-$19.31$\pm$0.03 &  $-$19.24$\pm$0.03 & $-$18.97$\pm$0.03 \\
Reindl et al. (2005) & $-$19.16$\pm$0.06 & $-$19.14$\pm$0.04 & $-$18.89$\pm$0.07 \\ 
Wang et al. (2005) &$-$19.27$\pm$0.09 &$-$19.20$\pm$0.09 &$-$18.86$\pm$0.09 \\ \hline
Average Values  & $-$19.28$\pm$0.08 & $-$19.20$\pm$0.05 & $-$18.90$\pm$0.06 \\ \hline \hline
Distance (Tab. \ref{SN_par1}) & $-$19.39$\pm$0.33 & $-$19.30$\pm$0.33&  $-$19.00$\pm$0.33\\ \hline
\end{tabular}
\\
$^\ddag$ Not considered in the computation of the average absolute magnitudes
\end{table}

An alternative method was recently proposed by \cite{wang05}, who 
introduced a new parameter, the 
intrinsic B--V colour 12 days after maximum light ($\Delta C_{12}$),
which is correlated with the absolute magnitude via the empirical 
formula: 

\begin{equation} \label{eq5}
M_{\lambda,max} = M_0 + R \times \Delta C_{12}
\end{equation}

where the parameter $\Delta C_{12}$ is found to be 0.354 $\pm$ 0.022 using
$\Delta m_{15}(B)$ = 1.12 and the relation of Wang et al.:

\begin{equation} \label{eq6}
\Delta C_{12} = 0.347 + 0.401\times D_{15}^B - 0.875 \times (D_{15}^B)^{2}
 + 2.44 \times (D_{15}^B)^{3}
\end{equation}

where $D_{15}^B$ = $\Delta m_{15}(B)$ -- 1.1.
The values of the parameters $M_0$ and $R$ are reported in Tab. 2 of \cite{wang05}. 
This provides $M_{B,max}$ = $-$19.27 $\pm$ 0.09. 
Estimates for the $V$ and $I$ band are also reported in Tab. \ref{abs_max}.

Averaging the calibrated absolute magnitudes obtained applying different
methods (those obtained with the older relations of
\cite{phil99} were not considered), the following estimates have been obtained for SN~2005cf:
$M_{B,max}$ = $-$19.28$\pm$0.08,
$M_{V,max}$ = $-$19.20$\pm$0.05 and $M_{I,max}$ = $-$18.90$\pm$0.06,
where the errors are the standard deviations of the available
estimates (see Tab. \ref{abs_max}). 
From the average absolute peak magnitudes, we obtain the following 
reddening--corrected colours:
$B-V$ = $-$0.08 and $V-I$ = $-$0.30, which are well consistent with 
those obtained from the direct measurements in Tab. \ref{SN_par1}, being $-$0.09 
and $-$0.30 respectively.

Another interesting parameter is the  rise time $t_r$ in the $B$ band,
i.e. the time spent by the SN from the explosion to the $B$ band maximum.  
A first attempt to estimate this parameter was performed by \citet{psk84},
who found it to be related to the post--maximum decay
rate $\beta$, closely related to the $\Delta m_{15}(B)$,
via the relation:

\begin{equation} \label{eq7}
t_r = 13 + 0.7 \beta
\end{equation}

For SN~2005cf $\beta$ is estimated to be about 7.47 mag/100$^{d}$,
setting the explosion epoch $t_r \sim$ 18.2 days before the $B$ band maximum. 
Another more recent method was suggested by \citet{ries99}.
In a first approximation, very young SNe Ia 
are homologously expanding fireballs, 
where the luminosity is proportional to the square of time since
explosion. \citet{ries99} derived $t_r$ from the relation:
\begin{equation} \label{eq8}
L(t) = A \times (t + t_r)^2
\end{equation}

where $t$ is the elapsed time relative to the maximum and
$A$ is a parameter describing the raising rate. Using very early
photometric data in the $R$ band (including the earliest unfiltered
measurements from IAU Circ. 8535, and considering data until $\sim$9 days before maximum),
we find that SN~2005cf exploded 19.2 days before the $R$ band maximum
(JD = 2453515.4). This corresponds to a rise time to maximum in the
$B$ band  $t_r$ $\approx$ 18.6$\pm$0.4 days,
not far from that derived applying Eq. \ref{eq7},
but slightly shorter than the 19.5$\pm$0.2 days
found by \citet{ries99} for an object with $\Delta m_{15}(B) \approx$ 1.1.
However,  the value of $t_r$ obtained for SN~2005cf is in good agreement with
that derived by \citet{ben04} for the similar declining SN 2002bo 
($t_r$ = 17.9$\pm$0.5)\footnote{After submission of our paper, a preprint was posted 
\protect\citep{con06} with estimates of the rise times for a sample of 73 SNe Ia. The
average value for low redshift SNe is 19.58 days, similar to the estimate of 
\protect\cite{ries99}, and somewhat higher that our estimate for 2005cf}.

\section{Light curve models and $^{56}$Ni Mass Estimate} \label{sect-model}

\begin{figure}   
\resizebox{\hsize}{!}{
 \includegraphics{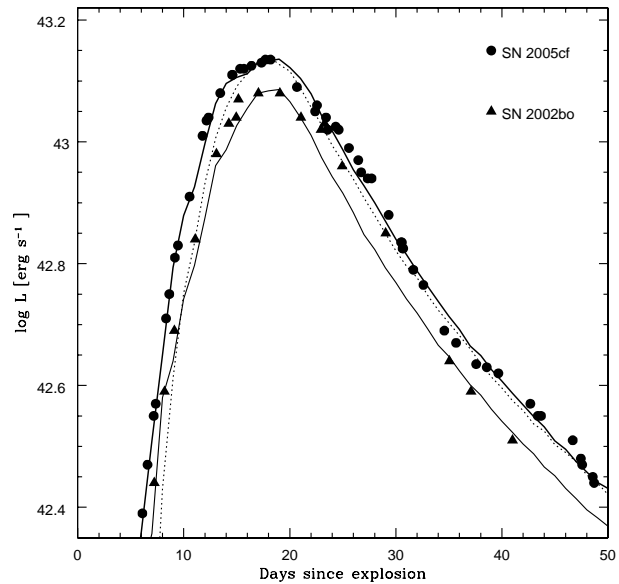}}
   \caption{Comparison between the observed bolometric light curves of
 SNe~2005cf (circles) and 2002bo (triangles), and the
 bolometric light curve models obtained using the W7 density
distribution and the abundances derived from the spectral analysis 
of Stehle et al. (2005). The dotted curve is a W7--based model with 0.7M$_\odot$ of $^{56}$Ni.
The two solid curves are a models obtained starting from Ni08-10$\%$ but increasing  the NSE abundances at 
high velocities as in Stehle et al. (2005): the thin lower curve is the model used for SN~2002bo, while
the thick higher is the model rescaled to the luminosity of
 SN~2005cf.} 
 \label{model}
\end{figure}

An useful tool to estimate the properties of a SN is the modelling of its 
bolometric light curve. 
The bolometric light curves of our SN sample were
computed applying the UV and NIR corrections from \citet{sunt96} to the observed quasi--bolometric light curves
of Fig. \ref{fig:bolometric}. 
The bolometric light curves of SNe 2005cf, 1992al and 2001el
are identical, while that of SN~2002bo is fainter (see also Fig. \ref{fig:bolometric} for a comparison).
In Fig. \ref{model} the bolometric light curves of SNe~2005cf and 2002bo
(those of SNe 1992al and 2001el are not shown) 
are compared with the models described below. The masses of the different components of the ejecta adopted in 
the models are reported in Tab. \ref{mod_par}.

We used a gray Monte Carlo light curve code \citep{maz01} to reproduce the bolometric
light curve and derive the properties of the ejecta. The code
computes the transport of the $\gamma$--rays and the positrons emitted by the
decay \Nifs\ $\rightarrow$ \Cofs\ $\rightarrow$ \Fefs, and then the transport of
the optical photons generated by the deposition of the energy carried by the
$\gamma$--rays and the positrons in the expanding SN ejecta. 
We assume that the ejecta mass is 1.4M$_\odot$, and that the density--velocity
distribution is described by the W7 model \citep{ken84}.
Compared to a model with 0.7M$_\odot$ derived with the W7 density
and abundance distributions (dotted curve in Fig.\ref{model}), the light curve 
of SN~2005cf (\DeltaB\ = 1.12, \DeltaV\ $\sim$ 0.6, \DeltaBol\ $\approx$ 0.9) is broader.

A factor that could make a light curve broad for its luminosity is the relative
content of ($^{58}$Ni+$^{54}$Fe) / $^{56}$Ni \citep{maz06}. For the
same total mass of material in nuclear statistical equilibrium (NSE),
a higher ($^{58}$Ni+$^{54}$Fe) / $^{56}$Ni ratio produces a dimmer light curve 
that has a comparable width. We find that the light curve of SN~2005cf is better 
fitted by models where the ratio is larger than in W7. 
In particular, in order to match the bolometric light curve of SN~2005cf, 
we start from a model where the $^{56}$Ni mass is intially rescaled to 0.8M$_\odot$, 
but then 10$\%$ of $^{56}$Ni is replaced with non--radioactive $^{58}$Ni and $^{54}$Fe 
(model Ni08--10$\%$, see \citet{maz06}). Such a
high ratio of stable versus radioactive Fe--group elements
($\sim$50$\%$) is unlikely for W7--like ignition conditions,
but is not excluded for higher ignition densities and/or
higher--than--solar metallicities \citep{friz05}.
\begin{table}
\caption{Parameters adopter in the light curve models shown in Fig. \ref{model}.}
\centering
\label{mod_par}
\begin{tabular}{ccccc}
\hline\hline
 Model  & ID curve  & M($^{56}$Ni) & M($^{58}$Ni+$^{54}$Fe) & M(IME+CO) \\ \hline
Ni07 & dotted & 0.7M$_\odot$ & 0.3M$_\odot$ & 0.4M$_\odot$ \\
Ni08--10$\%$ & $\ddag$ & 0.72M$_\odot$ & 0.38M$_\odot$ & 0.3M$_\odot$ \\ 
SN 2002bo & lower solid & 0.52M$_\odot$ & 0.36M$_\odot$ & 0.5M$_\odot$ \\
SN 2005cf & higher solid & 0.7M$_\odot$ & 0.4M$_\odot$ & 0.3M$_\odot$ \\ \hline 
\end{tabular}

$\ddag$ This model \citep[see][]{maz06} is not shown in Fig. \ref{model}
\end{table}

The model described above (not shown in Fig. \ref{model}) has M($^{56}$Ni) $\sim$ 0.72M$_\odot$ and
total M(NSE) $\sim$ 1.1M$_\odot$ (see Tab. \ref{mod_par}), but produces a luminosity peak that is too bright.  
In the case of SN~2002bo, the fast rise of
the light curve could be reproduced adopting the $^{56}$Ni distribution derived from
fitting a time sequence of spectra \citep[Fig. \ref{model}, thin lower solid curve]{steh05}. 
That distribution reached higher velocities than W7. 
In that model, M($^{56}$Ni) $\sim$ 0.52M$_\odot$ and M(NSE) $\sim$ 0.9M$_\odot$ were adopted.
Now we rescale our Ni08--10$\%$ model to the abundance 
distribution of the model used to fit SN~2002bo, although we cannot justify this with
spectroscopic results. The resulting model, with M($^{56}$Ni) $\sim$ 0.7M$_\odot$ 
and a mass of NSE elements of about 1.1M$_\odot$, fits the light curve of SN~2005cf quite well
(Fig. \ref{model}, thick higher solid curve). In total, 1.1M$_\odot$
are burned to NSE, and only 0.3M$_\odot$ are 
intermediate--mass elements (IME) or unburned material (CO). 
Similar values, in particolar an ejected $^{56}$Ni mass of about
0.7M$_\odot$, are also obtained for SNe~1992al and 2001el.
Spectroscopic models will be necessary to refine these estimates. 

\section{Summary} \label{sect-disc}

Extensive optical photometric observations of the nearby Type Ia SN~2005cf obtained by the
ESC are presented. The observations span a period of about 100 days, 
from $-$12 until +87 days from the $B$ band maximum.

Being a standard, normally--declining SN Ia, with a reddening corrected $\Delta
m_{15}(B)$ = 1.12, its light curves well match those of 
SNe~1992al and 2001el in optical bands. 
SN~2005cf can be considered a good template, having been discovered a short
time after the explosion and being densely sampled.
Despite some uncertainty in the distance of the host galaxy, 
its absolute magnitude at maximum (M$_B = -$19.39$\pm$0.33) is close to those
of SNe~1992al and 2001el, but SN~2005cf is probably intrinsically 
brighter than the similarly--declining SN~2002bo.
The colour evolution of SNe Ia in the $\Delta m_{15}(B)$ range
1.1--1.2 appears to be rather homogeneous.

The rise time of SN~2005cf to the $B$ band maximum is computed to be
18.6$\pm$0.4
days, slightly shorter than expected for a SN with such $\Delta m_{15}(B)$. 

Finally,  the bolometric light curve
modelling indicates an ejected $^{56}$Ni mass of about 0.7\Msun,
which is close to the average value of the $^{56}$Ni mass
distribution observed in normal SNe Ia
\citep[0.4--1.1 M$_\odot$, ][]{capp97}.

Spectroscopic data that will be presented in a forthcoming
paper \citep{gara06} will provide further information
about the properties of this normal object, and
the degree of homogeneity among the mid--declining SNe Ia.

\section*{Acknowledgments}
This work has been supported by the European Union's Human
Potential Programme ``The Physics of Type Ia Supernovae'', 
under contract HPRN-CT-2002-00303.
A.G. and V.S. would also like to thank the
G\"{o}ran Gustafsson Foundation for financial support.\\
This paper is based on observations collected at the
Centro Astron\'omico Hispano Alem\'an (Calar Alto, Spain), Siding Spring Observatory (Australia), 
Asiago Observatory (Italy), Telescopio Nazionale Galileo, Nordic Optical Telescope
and  Mercator Telescope (La Palma, Spain), ESO/MPI 2.2m Telescope (La
Silla, Chile), 2m Himalayan Chandra Telescope of the Indian
Astronomical Observatory (Hanle, India)\\
We thank the resident astronomers of the Telescopio Nazionale Galileo, the
Mercator Telescope, the ESO/MPI 2.2m Telescope
and the 2.2m and 3.5m telescopes in Calar Alto for performing
the follow--up observations of SN~2005cf.
We thank Thomas Augusteijn, Eija Laurikainen, Karri Muinonen and Pasi
Hakala for giving up part of their time at the Nordic Optical Telescope
(NOT), and Jyri N\"ar\"anen, Thomas Augusteijn, Heiki Salo, Panu Muhli,
Tapio Pursimo, Kalle Torstensson and Danka Parafcz for performing the
observations. Observations on Aug 15 were performed by the Olesja Smirnova and
Are Vidar Hansen as remote observations with the NOT at the Nordic
Baltic Research School: ``Looking Inside Stars'', which took place at
Moletai Observatory, Lithuania, August 7-21, 2005.
We are also grateful to P. Sackett
for the help in observing SN~2005cf from Siding Spring.\\
This research has made use of the NASA/IPAC Extragalactic
Database (NED) which is operated by the Jet Propulsion Laboratory,
California Institute of Technology, under contract with the National
Aeronautics and Space Administration. We also made use of the Lyon-Meudon
Extragalactic Database (LEDA), supplied by the LEDA team at
the Centre de Recherche Astronomique de Lyon, Observatoire de Lyon.\\

\bibliographystyle{mn2e}
\bibliography{biblio}

\begin{thebibliography}{}
\expandafter\ifx\csname natexlab\endcsname\relax\def\natexlab#1{#1}\fi

\bibitem[Altavilla et al., 2004]{alt04} Altavilla, G.
et al., 2004, MNRAS, 349, 1344



\bibitem[Astier et al., 2006]{asti06} Astier, P., et al, 2006
A\&A, 447, 31

\bibitem[Benetti et al., 2004]{ben04}
Benetti, S. et al., 2004, MNRAS, 348, 261

\bibitem[Benetti et al., 2005]{ben05}
Benetti, S. et al., 2005, ApJ, 623, 1011

\bibitem[Bessell, 1990]{bess90} Bessell, M. S.,
1990, PASP, 102, 1181



\bibitem[Cappellaro et al., 1997]{capp97} Cappellaro, E., Mazzali,
P. A., Benetti, S., Danziger, I. J., Turatto, M., della Valle, M.,
1997, A\&A, 328, 203


\bibitem[Conley et al., 2006]{con06} Conley, A., et al., 2006, AJ, 132, 1707

\bibitem[Contardo et al., 2000]{con00}
Contardo, G., Leibundgut, B., Vacca, W. D., 2000, A\&A, 359, 876

\bibitem[Della Valle \& Livio, 1994]{mdv94}  Della Valle, M., Livio, M.,
1994, ApJ, 423, L31 

\bibitem[Della Valle et al., 2005]{mdv05} Della Valle, M., Panagia, N.,
Padovani, P., Cappellaro, E., Mannucci, F., Turatto, M., 2005, ApJ,
629, 750

\bibitem[Elias--Rosa et al., 2006]{nancy06} Elias-Rosa, N. et al., 
2006, MNRAS, 369, 1880

\bibitem[Fouqu\'e et al., 2001]{fou01} Fouqu\'e, P., Solanes, J. M., 
Sanchis, T., Balkowski, C., 2001, A\&A, 375, 770

\bibitem[Garavini et al., 2007]{gara06} Garavini, G. et al., 2007, A$\&$A, in press

\bibitem[Hamuy et al., 1994]{ham94} Hamuy, M., Suntzeff, N. B.,
Heathcote, S. R., Walker, A. R., Gigoux, P., Phillips, M. M., 1994,
PASP, 106, 566

\bibitem[Hamuy et al., 1996]{hamu96} Hamuy, M. et al., 1996, AJ, 112, 2408

\bibitem[Hachinger et al., 2006]{hach06} Hachinger, S., Mazzali,
P. A., Benetti, S., 2006, MNRAS, 370, 299

\bibitem[Hopp \& Fern\`andez, 2002]{atm_caha} Hopp, U., Fern\`andez,
M., 2002, International Report 17/10/2002

\bibitem[Hunt et al., 1998]{hunt98} Hunt, L. K., Mannucci, F., Testi,
L., Migliorini, S., Stanga, R. M., Baffa, C., Lisi, F., Vanzi, L., 1998, AJ, 115, 2594

\bibitem[Jha et al., 2006]{jha06} Jha, S. et al., 2006, AJ, 131, 527

\bibitem[Kasen, 2006]{kas06} Kasen, D., 2006, ApJ, 649, 939

\bibitem[King, 1985]{atm_lapalma} King, D. L., 1985, RGO/La Palma
technical note No. 31

\bibitem[Kotak et al., 2005]{kot05} Kotak, R.
et al., 2005, A$\&$A, 436, 1021


\bibitem[Kraan--Korteweg, 1986]{kran86} Kraan--Korteweg, R. C., 1986,
A\&AS, 66, 255

\bibitem[Krisciunas et al., 2003]{kris03} Krisciunas, K., et al.,
2003, AJ, 125, 166

\bibitem[Krisciunas et al., 2004]{kris04} Krisciunas, K., et al.,
2004, AJ, 128, 3034

\bibitem[Landolt, 1992]{land92} Landolt, A. U., 1992,
AJ, 104, 340 

\bibitem[Mazzali et al., 2001]{maz01}
Mazzali, P. A., Nomoto, K., Cappellaro, E., Nakamura, T., Umeda, H., 
Iwamoto, K., 2001, ApJ, 547, 988

\bibitem[Mazzali et al., 2005]{maz05}
Mazzali, P. A. et al., 2005, ApJ, 623, 37

\bibitem[Mazzali \&  Podsiadlowski, 2006]{maz06}  Mazzali, P. A., 
Podsiadlowski, Ph., 2006, MNRAS, 369L, 19

\bibitem[Modjaz et al., 2005]{modj05} Modjaz, M., Kirshner, R.,
Challis, P., Berlind, P. 2005, IAU Circ. 8534, 3

\bibitem[Navasardyan et al., 2001]{nava01} Navasardyan, H., 
Petrosian, A. R., Turatto, M., Cappellaro, E., Boulesteix, J.,
2001, MNRAS, 328, 1181

\bibitem[Nomoto et al., 1984]{ken84} Nomoto, K., Thielemann, F.-K.,
Yokoi, K., 1984, ApJ, 286, 644 

\bibitem[Pastorello et al., 2007]{pasto06} Pastorello, A.  et al.,
2007, MNRAS, submitted

\bibitem[Perlmutter et al., 1997]{perl97}
Perlmutter, S.. et al., 1997, ApJ, 483, 565

\bibitem[Persson et al., 1998]{pers98} Persson, S. E. et al., 1998,
AJ, 116, 2475

\bibitem[Petrosian \& Turatto, 1995]{pet95} Petrosian, A. R., Turatto,
M., 1995, A\&A, 297, 49 

\bibitem[Phillips et al., 1993]{phil93}
Phillips, M. M., 1993, ApJ, 413, L105

\bibitem[Phillips et al., 1999]{phil99}
Phillips, M. M., Lira, P., Suntzeff, N. B., Schommer, R. A., 
Hamuy, M., Maza, J. 1999, AJ, 118, 1766

\bibitem[Pignata et al., 2004]{pig04} Pignata, G.
et al., 2004, MNRAS, 355, 178


\bibitem[Prieto et al., 2006]{pri06} Prieto, J. L., Rest, A.,
Suntzeff, N. B., 2006, ApJ, 647, 501

\bibitem[Pskovskii, 1984]{psk84} Pskovskii, Yu. P., 1984, Soviet
Astron., 28, 658 

\bibitem[Puckett et al., 2005]{puck05} Puckett, T., Langoussis, A., 
Chen, Y.-T., Hu, C.-P., Pugh, H., Li, W., Harris, B. 2005, IAU Circ. 8534, 1

\bibitem[Reindl et al., 2005]{rein05}
Reindl, B., Tammann, G. A., Sandage, A., Saha, A., 2005, ApJ, 624, 532


\bibitem[Riess et al., 1999a]{ries99a} Riess, A. G. et al., 1999a,
AJ, 117, 707

\bibitem[Riess et al., 1999b]{ries99} Riess, A. G. et al., 1999b,
AJ, 118, 2675

\bibitem[Roepke at al., 2006]{friz05} Roepke, F. K., Gieseler, M.,
Reinecke, M., Travaglio, C., Hillebrandt, W., 2006, A\&A, 453, 203


\bibitem[Schlegel et al., 1998]{schl98} Schlegel, D. J., 
Finkbeiner, D. P., Davis, M. 1998, ApJ, 500, 525

\bibitem[Silk, 1977]{silk77} Silk, J., 1977, A\&A, 59, 53

\bibitem[Smirnov \& Tsvetkov, 1981]{smi81} Smirnov, M. A., Tsvetkov,
D. Yu., 1981, Pis'ma Astr. Zh., 7, 154  

\bibitem[Somerville et al., 1997]{som97} Somerville, R. S., 
Davis, M., Primack, J. R., 1997, ApJ, 479, 616

\bibitem[Stanishev et al., 2007]{sta06} Stanishev, V.,
et al., 2007, A\&A, submitted

\bibitem[Stehle et al., 2005]{steh05} Stehle, M., Mazzali, P. A., 
Benetti, S., Hillebrandt, W., 2005, 360, 1231

\bibitem[Stritzinger et al., 2002]{max02}
Stritzinger, M. et al., 2002, AJ, 124, 2100

\bibitem[Stritzinger et al., 2005]{max05}
Stritzinger, M., Suntzeff, N. B., Hamuy, M., Challis, P., Demarco, R., 
Germany, L., Soderberg, A. M., 2005, PASP, 117, 810 

\bibitem[Suntzeff, 1996]{sunt96} Suntzeff, N. B., 1996 in McCray R.,
Wang Z. eds, Proc. IAU Colloquium 145, {\sl Supernovae and Supernova
Remnants}, Cambridge: University Press, P. 41 

\bibitem[Tammann \& Sandage, 1985]{tamm85} Tammann, G. A., Sandage,
A., 1985, ApJ, 294, 81

\bibitem[Terry et al., 2002]{terry02} Terry, J. N., Paturel, G.,
Ekholm, T., 2002, A\&A, 393, 57


\bibitem[Vorontsov-Velyaminov \& Arhipova, 1963]{voro63} 
Vorontsov-Velyaminov, B., Arhipova, V. P. 1963
{\it "Morphological Catalogue of Galaxies"}, Vol. 3, 
Moscow State University


\bibitem[Walker, 1987]{walker} Walker, G., 1987, Astronomical
Observations. Cambridge Univ. Press, Cambridge, p. 47

\bibitem[Wang et al., 2005]{wang05} Wang, X.,
Wang, L., Zhou, X., Lou, Y.-Q. Li, Z., 2005, ApJ, 620L, 87

\end{thebibliography}
\end{document}